\documentclass[fleqn,usenatbib]{mnras}

\usepackage{newtxtext,newtxmath}

\usepackage[T1]{fontenc}
\usepackage{subcaption}

\DeclareRobustCommand{\VAN}[3]{#2}
\let\VANthebibliography\thebibliography
\def\thebibliography{\DeclareRobustCommand{\VAN}[3]{##3}\VANthebibliography}

\usepackage{graphicx}	
\usepackage{newtxmath}	
\usepackage[dvipsnames]{xcolor}
\usepackage{enumitem}
\usepackage{bm}

\title[Relaxation of magnetically-confined mountains by cross-field transport]{Relaxation of magnetically-confined mountains on accreting neutron stars through cross-field mass transport}

\author[Brunet et al.]{
Ryan Brunet,$^{1,2}$\thanks{E-mail: rbrunet@student.unimelb.edu.au}
Andrew Melatos,$^{1,2}$
and Pedro H. B. Rossetto$^{3}$
\\
$^{1}$School of Physics, University of Melbourne, Parkville, VIC 3010, Australia\\
$^{2}$ARC Centre of Excellence for Gravitational Wave Discovery (OzGrav), University of Melbourne, Parkville, VIC 3010, Australia\\
$^{3}$ Escola de Artes, Ciências e Humanidades, Universidade de São Paulo, São Paulo, SP 03828-000, Brazil}

\date{Accepted XXX. Received YYY; in original form ZZZ}

\pubyear{2026}

\begin{document}
\label{firstpage}
\pagerange{\pageref{firstpage}--\pageref{lastpage}}
\maketitle

\begin{abstract}
Hydromagnetic instabilities modify the structure of a magnetically confined mountain on an accreting neutron star, once the accreted mass exceeds a critical value. Ideal magnetohydrodynamics and flux freezing break down, and mass diffuses across magnetic field lines locally, wherever instabilities are excited. Here a self-consistent, iterative, numerical scheme is used to evolve an axisymmetric magnetic mountain through a quasistatic sequence of Grad-Shafranov equilibria as a function of the accreted mass, $M_{\rm a}$, modified by instability-driven cross-field mass transport obeying the semi-analytic, Kulsrud-Sunyaev recipe. The results are compared to an artificially stabilised mountain, in which flux freezing does not break down, and there is no cross-field mass transport. It is shown that cross-field mass transport prevents instabilities from demolishing the mountain. Instead, the mass-flux distribution adjusts locally to nullify the instabilities and preserve a nonzero mass quadrupole moment indefinitely in the absence of ohmic dissipation.
\end{abstract}

\begin{keywords}
accretion -- instabilities -- magnetic fields -- MHD -- stars: neutron -- turbulence
\end{keywords}


\section{Introduction}{\label{sec:1}}
Rapidly rotating, non-axisymmetric neutron stars are canonical sources of continuous gravitational waves \citep{2015PASA...32...34L, 2018ASSL..457..673G}. Non-axisymmetries arise from several mechanisms in both isolated and accreting systems, including deformations due to the strong internal magnetic field \citep{1996A&A...312..675B,2002PhRvD..66h4025C,2008MNRAS.385..531H, 2010MNRAS.406.2540C}, temperature gradients in the crust leading to elastic strain \citep{1998ApJ...501L..89B,2000AIPC..523...65U,2006MNRAS.373.1423H, 2020MNRAS.494.2839O,2021MNRAS.507..116G,2022MNRAS.514.1628K}, and superfluid vortex pinning in the stellar interior \citep{2002CQGra..19.1255J,2010MNRAS.402.2503J,2015ApJ...807..132M,2022Univ....8..619H,2024MNRAS.528.1360C}. In accreting systems such as low mass X-ray binaries, non-axisymmetry is also generated, when accreted material is funnelled onto and confined at the magnetic pole by the Lorentz force of the stellar magnetic field, forming a localised `magnetic mountain'. In this scenario, the accreted mass accumulates in a column over the polar cap, distorting the magnetic field as the hydrostatic pressure balances the Lorentz force \citep{1983A&A...128..369H, 1998ApJ...496..915B, 2001ApJ...553..788L,2001PASA...18..421M,2004MNRAS.351..569P, 2008MNRAS.386.1294V, 2009MNRAS.395.1985V, 2011MNRAS.417.2696P, 2019MNRAS.484.1079S}.
The magnetic field is effectively buried under the accreted overburden, reducing the global magnetic dipole moment $|{\bf m}|$ as the accreted mass $M_{\rm a}$ increases, in line with observations \citep{1986ApJ...305..235T, 1991PhR...203....1B,1995xrbi.nasa..495B, 1995A&A...297L..41V}. The mass ellipticity \(\epsilon\) also increases, as $M_{\rm a}$ increases, amounting to a mountain with height $\epsilon R_\ast \lesssim 0.1\,\rm m$ for theoretical estimates  $\epsilon\lesssim10^{-5}$ obtained from several studies \citep{2005ApJ...623.1044M,2009MNRAS.395.1985V, 2011MNRAS.417.2696P,2025ApJ...979...10R}. Gravitational radiation reaction may explain why neutron stars are observed to spin at frequencies significantly lower than the centrifugal break-up frequency of approximately 1\,kHz \citep{1998ApJ...501L..89B, 2003Natur.424...42C,2005MNRAS.361.1153A}.

A polar magnetic mountain evolves away from its equilibrium configuration for various reasons. On time-scales longer than the Alfv\'{e}n time, it relaxes due to a combination of Ohmic dissipation \citep{2009MNRAS.395.1985V}, sinking of accreted material into the crust \citep{2002MNRAS.332..933C,2010MNRAS.402.1099W}, and thermal conduction \citep{2019MNRAS.484.1079S}. The structure is also modified by the Hall effect \citep{2004ApJ...609..999C}, the equation of state (EOS) \citep{2011MNRAS.417.2696P}, and the presence of neutron superfluidity and proton superconductivity in the core \citep{2009MNRAS.395.2162L,2012MNRAS.419..732L,2012MNRAS.420.1263G,2021PASA...38...43S}. 
On timescales comparable to the Alfv\'{e}n timescale, it is expected that the highly-distorted equilibrium is prone to magnetohydrodynamic (MHD) instabilities, such as the ballooning mode instability \citep{RevModPhys.54.801,2001ApJ...553..788L}. \citet{2007MNRAS.376..609P} and \cite{2008MNRAS.386.1294V} investigated axisymmetric and non-axisymmetric instability modes and found that an unstable mountain oscillates in a superposition of acoustic and Alfv\'{e}n modes but remains intact --- with mass ellipticity reduced by $\sim$30 per cent --- due to the stabilising effect of magnetic line-tying at the surface. In a complementary suite of time-dependent MHD simulations with different initial and boundary conditions, \citet{2013MNRAS.430.1976M, 2013MNRAS.435..718M} observed that mountains are unstable to interchange instabilities above a threshold mass in two and three dimensions.

Importantly, all the numerical calculations referenced above face a daunting challenge: the mountain dynamics span a wide --- and indeed unresolvable --- range of time-scales (from the short Alfv\'{e}n crossing time to the long accretion time) and length-scales (from the microscopic wavelength of the fastest-growing unstable mode to the macroscopic size of the magnetosphere). To circumvent the limitations imposed by numerical resolution, \citet{2020JPlPh..86f9002K} proposed a semi-analytic picture based upon the ideal Schwarzschild instability \citep{schwarzschild_1958,newcomb1961convective}, which produces a cascade of eddies extending down to the resistive scale, where the flux-freezing constraint is violated, and mass diffuses across magnetic flux surfaces to maintain marginal stability on every field line.

In this paper, we develop the semi-analytic picture of cross-field mass transport pioneered by \citet{2020JPlPh..86f9002K} to study how MHD instabilities occurring on the fast Alfv\'{e}n time-scale modify the long-term evolution of a polar magnetic mountain on the slow accretion time-scale $M_{\rm a}/\dot{M}_{\rm a}$. Specifically, \citet{2020JPlPh..86f9002K} argued that a mountain exists in a state of marginal stability over the long term: local MHD instabilities are triggered by the incremental addition of accreted mass, flux freezing breaks down locally, matter diffuses across field lines in the unstable region, equilibrium is restored temporarily, and the cycle repeats. This subtle process cannot be studied numerically from first principles with realistic time-dependent MHD simulations, because the separation between the Alfv\'{e}n and accretion time-scales is too great. Instead we track the unstable dynamics in a step-wise fashion through a quasistatic sequence of equilibria by combining a well-tested Grad-Shafranov solver \citep{2004MNRAS.351..569P,2008MNRAS.386.1294V} with the semi-analytic cross-field mass transport scheme proposed by \citet{2020JPlPh..86f9002K}. By iterating between transporting mass across unstable flux surfaces and recalculating the hydromagnetic equilibrium, we confirm explicitly that the mountain self-adjusts to remain marginally stable on every flux surface on the slow accretion time-scale $M_{\rm a} / \dot{M}_{\rm a}$. We study a range of accreted masses, and quantify the resulting trends of \(|{\bf m}|\) and \(\epsilon\) versus \(M_{\rm a}\). The paper is structured as follows. In Section \ref{sec:2} we review the theory of MHD instabilities on accreting neutron stars, with a special focus on magnetic mountains. In Section \ref{sec:3}, we describe the mathematical framework used to calculate hydromagnetic equilibria on accreting neutron stars, the instability criterion and cross-field mass transport recipe developed by \citet{2020JPlPh..86f9002K}, and their combination in an iterative numerical scheme. In Section~\ref{sec:4} we show that the semi-analytic recipe nullifies unstable regions, so that a mountain remains largely intact for \(M_{\rm a} \lesssim 10^{-5} M_\odot\). In Section~\ref{sec:astro_observables}, we calculate the corresponding magnetic dipole moments and ellipticities and discuss briefly their astrophysical implications, e.g.\ for gravitational wave astronomy \citep{2023LRR....26....3R}. Technical details of the numerical scheme are recorded in the appendices for the sake of reproducibility.

\section{Mountain Stability}\label{sec:2}
Instabilities are an endemic feature of magnetically confined plasmas in space and terrestrial laboratories \citep{1986islp.book.....M} and are a key stumbling block in developing thermonuclear fusion devices (e.g. tokamaks; \citealp{2016ippc.book.....C}) as a practical source of grid-scale electrical power. Taken at face value, one would expect polar magnetic mountains on accreting neutron stars to be unstable to the magnetic Rayleigh-Taylor instability and its variants: the accreted material buries the star's polar magnetic field, bending the field lines into an ``equatorial tutu'', such that lighter (strongly magnetised) fluid supports heavier (weakly magnetised) fluid against gravity. Counter-intuitively, however, many studies find that polar magnetic mountains are more stable than one expects naively. Instabilities do occur, but they saturate due to various mechanisms and do not disrupt the mountain completely \citep{2001ApJ...553..788L, 2006ApJ...652..597P,2007MNRAS.376..609P, 2008MNRAS.386.1294V, 2009MNRAS.395.1985V, 2013MNRAS.430.1976M,2013MNRAS.435..718M,2020JPlPh..86f9002K}. 

In Section \ref{sec:mhd_instabilities}, we review previous studies of magnetic mountain stability, in order to set the context for the present work. The reviewed studies uncover several subtle pieces of physics, which control mountain stability and are essential for interpreting the calculations in the present paper. We focus on local ideal MHD instabilities. Resistive processes such as Ohmic diffusion \citep{2009MNRAS.395.1985V}, thermal transport \citep{2019MNRAS.484.1079S}, and large-scale sinking \citep{2002MNRAS.332..933C, 2010MNRAS.402.1099W, 2020MNRAS.499.3243S} are postponed to future work. In Section \ref{sec:unstable cf transport}, we explain how the semi-analytic picture of cross-field mass transport caused by local ideal MHD instabilities, proposed by \citet{2020JPlPh..86f9002K} and implemented in this paper, fits together with previous work.

\subsection{MHD instabilities}{\label{sec:mhd_instabilities}}
The standard treatment of stability in astrophysical plasmas begins with the MHD energy principle \citep{1958RSPSA.244...17B}. A hydromagnetic equilibrium with mass density $\rho({\bf x})$, pressure $p({\bf x})$, gravitational potential $\Phi({\bf x})$, and magnetic field strength ${\bf B}({\bf x})$ is unstable, if there exists a complex Lagrangian displacement $\bm{\xi}({\bf x})$ whose associated change in potential energy $\delta W$ is negative, with \citep{1993noma.book.....B, 2004prma.book.....G}
\begin{align}\label{eq:dW_01}
    \delta W = &\ \frac{1}{2}\int dV\,\left[|\mathbf{Q}|^2 + \gamma p |\nabla\cdot\bm{\xi}|^2 + (\mathbf{\xi^*}\cdot\nabla p)\nabla\cdot\bm{\xi}+\mathbf{j}\cdot\bm{\xi}\times\mathbf{Q} \right. \nonumber\\
    & \left.-(\mathbf{\xi^*}\cdot\nabla\Phi)\nabla\cdot(\rho\bm{\xi}) \vphantom{\gamma p |\nabla\cdot\bm{\xi}|^2 + |\mathbf{Q}|^2 + (\mathbf{\xi^*}\cdot\nabla p)\nabla\cdot\bm{\xi}+\mathbf{j}\cdot\bm{\xi}\times\mathbf{Q}}\right].
\end{align}
In equation \eqref{eq:dW_01}, which involves an integral over the system's volume, we define \(\mathbf{Q} = \nabla\times(\bm{\xi}\times\mathbf{B})\), the adiabatic index \(\gamma\), and the current density \(\mathbf{j} = \nabla\times\mathbf{B}/\mu_0\) (SI units). The symbol \(|\dots|\) denotes the complex modulus. The first two terms are positive definite (stabilising) and represent the magnetic and acoustic energy respectively. The final three terms may be negative (destabilising) and cause pressure-driven (e.g. ballooning) instabilities, current-driven instabilities, and gravitational (e.g. Rayleigh-Taylor, Kruskal-Schwarzschild, Parker) instabilities respectively \citep{2004prma.book.....G} .

Accreted mountains on neutron stars typically have regions with \(\bf{B}\perp\nabla \Phi\) and are therefore susceptible to gravitational Kruskal-Schwarzschild interchange modes \citep{1954RSPSA.223..348K,newcomb1961convective,1974ApJ...192...37M,1987GApFD..39...65H, 2001ApJ...553..788L,2004MNRAS.351..569P}. \citet{2001ApJ...553..788L} applied \eqref{eq:dW_01} to determine the onset of gravitational instabilities due to the overpressure \(\Delta p\) created by an accreted mountain. Line-tying of the magnetic field at the surface eliminates interchange modes \citep{2008MNRAS.386.1294V}, but other ballooning modes such as the undular Parker instability may occur for $M_{\rm a} \gtrsim 10^{-10} M_{\odot}$ \citep{2001ApJ...553..788L}. Back reaction from the compressed equatorial magnetic field stabilises the system \citep{2004MNRAS.351..569P}.%
Detailed numerical investigations of the stability of magnetic mountains were performed by \citet{2007MNRAS.376..609P} and \citet{2008MNRAS.386.1294V} in two and three dimensions respectively. Both analyses start from a Grad-Shafranov equilibrium, whose flux function is determined uniquely by connecting to the pre-accretion magnetic field through flux-freezing (as in the present paper), unlike other analyses, where the flux function is arbitrary. This is important, because subtle features of the magnetic geometry stemming from the flux-freezing constraint can make all the difference between stability and instability when comparing two otherwise similar field configurations \citep{2004prma.book.....G}. In two dimensions, \citet{2007MNRAS.376..609P} found that isothermal magnetic mountains are marginally stable for $M_{\rm a} \lesssim 6 \times 10^{-4} M_\odot$; that is, the mountain wobbles due to acoustic and Alfv\'{e}n oscillations but it is not disrupted. Magnetic bubbles (i.e.\, closed magnetic loops) form for $M_{\rm a} \gtrsim 10^{-4} M_\odot$ but they are not caused by an unstable perturbation, whose amplitude grows exponentially. Rather, they are caused by a loss of equilibrium at a bifurcation point, where the quasistatic sequence of two-dimensional Grad-Shafranov equilibria terminates above a critical $M_{\rm a}$ value, and a different sequence of equilibria (e.g., non-axisymmetric ones) takes over \citep{1989ApJ...345.1034K}. In three dimensions, \citet{2008MNRAS.386.1294V} found that an initially axisymmetric isothermal mountain is unstable to the toroidal undular sub-mode of the Parker instability for $M_{\rm a} \gtrsim 1.2 \times 10^{-4} M_\odot$, but the instability saturates without disrupting the mountain. Instead, the system adjusts to a new, non-axisymmetric, mountain-like equilibrium, whose mass quadrupole moment is $\approx 30$ per cent smaller than before the instability (after spreading equatorward). Magnetic line tying at the surface plays a crucial stabilising role; it suppresses the interchange sub-mode, leaving the undular sub-mode -- which compresses plasma along field lines -- as the only unstable mode. The findings are consistent with the stability properties of solar prominences \citep{1984smh..book.....P} and tokamaks \citep{Lifshits1989-ha, 2004prma.book.....G} in two and three dimensions \citep{1992PASJ...44..167M}. \citet{2008MNRAS.386.1294V} also investigated systematically the possibility that the counter-intuitive nonlinear stability of the mountain may be a numerical artefact, e.g.\ due to inadequate spatial resolution. They concluded otherwise: the linear growth rate scales with the wavelength \(\lambda\) of the perturbation as \(\sim \lambda^{-1/2}\), suggesting that the dynamics are replicated at higher grid resolutions. \citet{2009MNRAS.395.1985V} investigated the effects of nonideal resistivity, finding no evidence of resistive ballooning or tearing modes on the Alfv\'{e}n time-scale \(\tau_{\mathrm{A}} = L\rho^{1/2}/B\), where \(L = (|\mathbf{B}|/|\nabla^2\mathbf{B}|)^{1/2}\) is a characteristic length scale, or the intermediate tearing mode time-scale \((\tau_{\mathrm{A}}\tau_{\mathrm{D}})^{1/2}\), where \(\tau_{\mathrm{D}} = L^2\sigma\) is the diffusion timescale and \(\sigma\) is the electrical conductivity, even after artificially raising the resistivity to encourage such modes. In combination with the line-tying boundary condition, finite resistivity enables magnetic reconnection to occur, smoothing toroidal field gradients and restoring non-axisymmetric field configurations back to axisymmetry on the slow Ohmic time-scale $\sim 10^7 \, {\rm yr} \sim M_{\rm a} / \dot{M}_{\rm a}$.

\cite{2013MNRAS.430.1976M, 2013MNRAS.435..718M} investigated the stability of axisymmetric magnetic mountains in two and three dimensions respectively using the state-of-the-art time-dependent MHD solver PLUTO \citep{2007jena.confR..96M}, by adding a random perturbation to the equilibrium density \citep{2013MNRAS.430.1976M} and velocity \citep{2013MNRAS.435..718M} fields, in order to trigger the growth of unstable modes. Their conclusions are broadly similar to those reviewed in the previous paragraph, viz.\ mountains are unstable above a threshold mass. Some subtle differences in the findings are traceable to the following differences in the set-up: (i) mass is strictly contained within the polar cap, rather than loaded smoothly over all field lines up to the equator; (ii) the equation of state is for a degenerate Fermi plasma; and (iii) the flux function defining the Grad-Shafranov equilibrium is specified as quadratic by fiat; it is not calculated self-consistently with reference to flux freezing and the pre-accretion state. \citet{2013MNRAS.430.1976M} investigated instabilities in two dimensions by adding a positive-definite random density field to the mountain equilibrium. The random field is added without adjusting the pressure to maintain equilibrium, so the additional mass falls under gravity, triggering magnetic Rayleigh-Taylor modes. The threshold perturbation amplitude for triggering the instability reduces, as the mountain height increases. Closed magnetic loops do not decay during  the simulation, and the instabilities persist for a range of grid resolutions. In three dimensions, \citet{2013MNRAS.435..718M} perturbed the equilibrium velocity field and observed similar behaviour: above a threshold perturbation amplitude, unstable toroidal modes are excited, generating radially elongated density streams.
\subsection{Local Schwarzschild instability leading to cross-field mass transport}\label{sec:unstable cf transport}
In response to the computational challenges posed by inadequate spatial resolution and a formidable dynamic range, discussed in Section~\ref{sec:mhd_instabilities}, \citet{2020JPlPh..86f9002K}  developed a semi-analytic picture of mountain relaxation in which a local, short-wavelength instability drives a turbulent cascade down to the resistive scale, as the instability evolves nonlinearly. At the resistive scale, mass migrates across flux surfaces in such a way as to throttle the instability and restore a state of marginal stability along every field line.

Several instabilities can viably play the role envisioned by \citet{2020JPlPh..86f9002K}. In this paper, we follow the latter authors and focus on the Schwarzschild instability, which operates similarly to the ballooning mode instability investigated by \citet{2001ApJ...553..788L} and the undular Parker sub-mode simulated by \citet{2008MNRAS.386.1294V}. It is a buoyancy-driven instability which occurs, when a component of the magnetic field is directed perpendicular to gravity, a situation which occurs, when the field is distorted by the accumulation of material in the polar mountain. 
Letting \(\xi\) denote the scalar fluid displacement perpendicular to the magnetic field, and upon defining unit vectors \(\mathbf{\hat{e}}_s\) along the magnetic field \(\bf B\), \(\mathbf{\hat{e}}_\phi\) in the azimuthal direction, and \(\mathbf{\hat{e}}_\xi\) orthogonal to both, equation \eqref{eq:dW_01} reduces to
\begin{equation}
    \delta W = \frac{1}{2}\int dV\,\left[\xi^2 P'\left(\frac{P'}{\Gamma P}  - \frac{\rho '}{\rho}\right) + \frac{1}{\mu_0}\left(B\frac{\partial \xi}{\partial s}\right)^2\right],
    \label{eq:dW_02}
\end{equation}
where \(P = p + B^2/8\pi\) is the total pressure, a prime indicates a derivative in the \(\xi\)-direction (i.e. normal to magnetic field lines), and we have \(\Gamma P = \gamma p - B^2/4\pi\), where \(\gamma\) denotes the adiabatic index. Hydrostatic equilibrium implies \(P' = -\rho g_\xi\), where \(g_\xi\) is the component of gravity in the \(\xi\)-direction (i.e.\ perpendicular to ${\bf B}$). Hence the first term in \eqref{eq:dW_02} becomes \(-\rho g_\xi(\Delta/C)\xi^2\), with \(C = \gamma(1 + \gamma\beta/2)\), and
\begin{align}
    \Delta &= \frac{p'}{p} - \frac{\gamma B'}{B}\nonumber\\
    & = \frac{d}{d\xi}\ln\left(\frac{p}{B^\gamma}\right).
\label{eq:delta_criterion_01}
\end{align}
The second term in \eqref{eq:dW_02} is positive-definite and corresponds to a tension force per unit volume, which must be overcome for the instability to grow. For \(\xi(s) = \cos(\pi\ell s)\), where \(s\) is the arc length along \(\bf B\), the second term in \eqref{eq:dW_02} has an average value equal to \citep{2020JPlPh..86f9002K}
\begin{equation}
    \left\langle\frac{1}{\mu_0}\left(B\frac{\partial \xi}{\partial s}\right)^2\right\rangle = \frac{\pi^2B^2\xi^2}{2\mu_0\ell^2}.
\end{equation}
Hence the instability condition \(\delta W\) evaluates to
\begin{equation}
    g_\xi(\Delta - \Delta_{\rm crit})> 0,
    \label{eq:delta_criterion_00}
\end{equation}
with
\begin{equation}
    \Delta_{\mathrm{crit}} = \frac{\pi^2 v^2_A C}{g_\xi\ell^2},
    \label{eq:delta_crit_01}
\end{equation}
where \(v_A = B/\sqrt{\mu_0\rho}\) is the Alfv\'{e}n speed. Equation~\eqref{eq:delta_criterion_00} implies that the Schwarzschild instability is triggered for \(g_\xi>0\) and \(\Delta > \Delta_{\rm crit}\), or \(g_\xi < 0\) and \(\Delta < \Delta_{\rm crit}\) in principle. However, only the former combination is observed in magnetic mountain equilibria \citep{2004MNRAS.351..569P, 2013MNRAS.430.1976M}. Equations~\eqref{eq:delta_criterion_00} and \eqref{eq:delta_crit_01} are reminiscent of convective instabilities, where the logarithmic gradient of the entropy \(\Delta = {d}\ln\left(p/\rho^{\gamma}\right)/d\xi\) determines the stability of a convective cell. A semi-analytic picture of the nonlinear development of the Schwarzschild instability, and its implications for cross-field mass transport \citep{2020JPlPh..86f9002K}, are reviewed in Section \ref{sec:3}.
\section{Hydromagnetic equilibria modified by cross-field mass transport}{\label{sec:3}}
In this section, we set out a semi-analytic mathematical framework for describing the physical picture studied in this paper, namely the evolution of a magnetic mountain through a quasistatic sequence of MHD equilibria, where the evolution is driven by the slow accretion of plasma onto polar magnetic flux surfaces (governed by ideal MHD) and the rapid, nonideal transport of plasma across magnetic flux surfaces, when the Schwarzschild instability is triggered locally. We present the Grad-Shafranov equation describing ideal-MHD equilibrium in Section \ref{subsec3_01:hydro equilibrium}, the field line integral describing ideal-MHD flux conservation in Section \ref{subsec:3_02:prescribing mfd}, and the semi-analytic recipe for cross-field mass transport proposed by \citet{2020JPlPh..86f9002K} in Section \ref{subsec: cfmf}. We justify the quasistatic approximation in Section \ref{subsec:quasistatic_approx} by estimating the relevant short and long time-scales. We outline its numerical implementation in Section \ref{sec:3_numerical_implementation}, with references to the appendices for details.
\subsection{Grad-Shafranov equation}\label{subsec3_01:hydro equilibrium}
The equations of nonideal MHD in SI units comprise the equation of mass conservation,
\begin{equation}
    \frac{\partial \rho}{\partial t} + \nabla\cdot(\rho\boldsymbol{v}) = 0,
    \label{eq:mass_cons_01}
\end{equation}
the equation of momentum conservation,
\begin{equation}
    \rho\frac{\partial\boldsymbol{v}}{\partial t} + \rho(\boldsymbol{v}\cdot\nabla)\boldsymbol{v} = - \nabla p -\rho\nabla\Phi + \frac{1}{\mu_0}(\nabla\times\boldsymbol{B})\times\boldsymbol{B},
    \label{eq:momentum_cons_01}
\end{equation}
and the induction equation, without the displacement current,
\begin{equation}
    \frac{\partial\boldsymbol{B}}{\partial t} - \nabla \times (\boldsymbol{v}\times\boldsymbol{B}) = \frac{1}{\mu_0\sigma}\nabla^2 \boldsymbol{B},
    \label{eq:induction_eq_01}
\end{equation}
supplemented by an equation of state, \(p = K\rho^\Gamma\). Here, \(\boldsymbol{B}, \rho, p, \Phi, \boldsymbol{v}\), and \(\sigma\) denote the magnetic field, mass density, pressure, gravitational potential, plasma bulk velocity, and electrical conductivity, respectively, and \(\mu_0 = 4\pi\times 10^{-7} \mathrm{N~A^{-2}}\) is the vacuum permeability. In the magnetostatic limit  \(\boldsymbol{v} = 0\) and \(\partial/\partial t = 0\), equation \eqref{eq:momentum_cons_01} reduces to
\begin{equation}
    0 = \nabla p + \rho\nabla\Phi - \frac{1}{\mu_0}(\nabla\times\boldsymbol{B})\times\boldsymbol{B}.
    \label{eq:momentum_cons_02}
\end{equation}

In this paper, we are interested in axisymmetric equilibria, where the magnetic field may be expressed in terms of a scalar flux function \(\psi(r,\theta)\), viz.
\begin{equation}
    \boldsymbol{B} = \frac{\nabla \psi \times \hat{\mathbf{e}}_\phi}{r\sin\theta},
    \label{eq:axisym_B_01}
\end{equation}
where we adopt spherical polar coordinates \((r,\theta,\phi)\), and \(\theta=0\) corresponds to the magnetic pole. Upon substituting \eqref{eq:axisym_B_01} into \eqref{eq:momentum_cons_02}, we arrive at
\begin{equation}
    0 = \nabla p + \rho\nabla\Phi + (\Delta_\ast\psi)\nabla\psi,
    \label{eq:GS_eqn_01}
\end{equation}
where the Grad-Shafranov operator \(\Delta_\ast\) in spherical polar coordinates is given by
\begin{equation}
    \Delta_\ast = \frac{1}{\mu_0r^2\sin^2\theta}\left[\frac{\partial^2}{\partial r^2} + \frac{\sin\theta}{r^2}\frac{\partial}{\partial\theta}\left(\frac{1}{\sin\theta}\frac{\partial}{\partial\theta}\right)\right].
    \label{eq:GS_operator}
\end{equation}
The field-aligned component of \eqref{eq:GS_eqn_01} can be solved exactly, yielding
\begin{equation}
    \int \frac{dp}{d\rho}\frac{d\rho}{\rho} = F(\psi) - (\Phi - \Phi_0),
    \label{eq:GS_eqn_02}
\end{equation}
where \(\Phi_0 = \Phi(R_\ast)\) is a reference potential, and \(F(\psi)\) is a scalar function of \(\psi\) determined by the accretion physics. For an isothermal equation of state \(p = c_s^2 \rho\), equation \eqref{eq:GS_eqn_02} reduces to
\begin{equation}
    p = F(\psi)\exp\left(-\frac{\Phi - \Phi_0}{c_s^2}\right),
    \label{eq:pressure_isothermal_exp_01}
\end{equation}
and \eqref{eq:GS_eqn_01} simplifies to the Grad-Shafranov equation
\begin{equation}
    \Delta_\ast\psi = -F^{\prime}(\psi)\exp\left(-\frac{\Phi - \Phi_0}{c_s^2}\right).
    \label{eq:GS_eqn_03}
\end{equation}

Equation \eqref{eq:GS_eqn_03} is an elliptic second-order partial differential equation which can be solved by standard methods given \(F(\psi)\) \citep{evans_pde_textbook}. In the neutron star magnetic mountain literature, one can either specify \(F(\psi)\) by fiat \citep{1983A&A...128..369H,1998ApJ...496..915B,2001ApJ...553..788L,2012MNRAS.419..732L,2012MNRAS.424..482L,2013MNRAS.430.1976M, 2013MNRAS.435..718M,2021PASA...38...43S, 2025MNRAS.541.3280Y} or compute it self-consistently via an integral constraint describing MHD flux freezing \citep{2004MNRAS.351..569P,2008MNRAS.386.1294V,2011MNRAS.417.2696P,2019MNRAS.484.1079S,2023MNRAS.526.2058R,2025ApJ...979...10R}. In this paper, we adopt the latter approach, which involves solving \eqref{eq:GS_eqn_03} and the integral constraint iteratively by a relaxation method, as discussed in Section \ref{subsec:3_02:prescribing mfd}.
\subsection{Mass-flux distribution}\label{subsec:3_02:prescribing mfd}
The mass-flux distribution \(dM/d\psi\) is defined as the mass contained between infinitesimally separated flux surfaces \(\psi\) and \(\psi + d\psi\). It is given by the field line integral
\begin{equation}
    \frac{dM}{d\psi} = 2\pi\int_{\mathcal{C}}ds\, r\sin\theta|\nabla\psi|^{-1}\rho\left[r(s),\theta(s)\right],
    \label{eq:dmdpsi_01}
\end{equation}
where \(\mathcal{C}\) describes the curve \(\psi\left[r(s),\theta(s)\right] = \psi\) parameterised by the arc length \(s\) along the intersection between the flux surface \(\psi\) and an arbitrary (due to axisymmetry) meridional plane.
Combining \eqref{eq:pressure_isothermal_exp_01} and the isothermal EOS \(p = c_s^2\rho\), equation \eqref{eq:dmdpsi_01} may be inverted to give
\begin{equation}
    F(\psi) = \frac{c_s^2}{2\pi}\frac{dM}{d\psi}\left\{\int_\mathcal{C}ds\, r\sin\theta|\nabla\psi|^{-1}e^{-(\Phi - \Phi_0)/c_s^2}\right\}^{-1}.
    \label{eq:Fpsi_01}
\end{equation}
One then differentiates \eqref{eq:Fpsi_01} with respect to \(\psi\) before substituting into \eqref{eq:GS_eqn_03}. A similar calculation applies for polytropic equations of state of the form \(p = K\rho^{\Gamma}\)\citep{2011MNRAS.417.2696P}, which will be analysed in a forthcoming paper. 

In ideal MHD, flux freezing guarantees that the mass between two flux surfaces equals the mass that is present there initially (before accretion) plus the mass that is loaded by accretion, which depends on how the flux surfaces connect to the accretion disk \citep{2004MNRAS.351..569P}. In this paper, the initial \(dM/d\psi\) is modified by cross-field mass transport according to the nonideal MHD recipe proposed by \citet{2020JPlPh..86f9002K}. We explain how the mass transport recipe is implemented in Section \ref{subsec: cfmf} and present the resulting formula for \(dM/d\psi\) in Section \ref{subsec:quasistatic_approx}.

We choose an initial \(dM/d\psi\) following the phenomenological form of previous authors, making the approximation \citep{2004MNRAS.351..569P, 2008MNRAS.386.1294V,2011MNRAS.417.2696P,2023MNRAS.526.2058R, 2025ApJ...979...10R}
\begin{equation}
    \frac{dM}{d\psi} = \frac{M_{\rm a}}{2\psi_{\rm a}}\frac{e^{-\psi/\psi_{\rm a}}}{1 - e^{-\psi_{\rm a}/\psi_\ast}}.
    \label{eq:dmdpsi_pm04}
\end{equation}
In \eqref{eq:dmdpsi_pm04}, \(M_{\rm a}\) is the total accreted mass, and \(\psi_\ast = \psi(R_\ast,\pi/2)\) defines the flux surface at the equator. As indicated by \eqref{eq:dmdpsi_pm04}, we allow accretion to  occur for $\psi_{\rm a} \leq \psi \leq \psi_\ast$ to avoid numerical issues from a sharp step-change in the density, improving the convergence of the solver.

Equations \eqref{eq:GS_eqn_03} and \eqref{eq:Fpsi_01} admit magnetic mountain solutions for $M_{\rm a} \leq M_{\rm c} \sim 10^{-5} M_\odot$ \citep{2004MNRAS.351..569P,2008MNRAS.386.1294V,2011MNRAS.417.2696P,2019MNRAS.484.1079S,2023MNRAS.526.2058R}. The quasistatic sequence of solutions terminates for $M_{\rm a} > M_{\rm c}$, where closed magnetic loops form \citep{1989ApJ...345.1034K}.

We emphasise that the polar mountain described by \eqref{eq:dmdpsi_pm04} is one of many plausible astrophysical solutions. A comprehensive study of alternative scenarios has been conducted recently by \citet{2025MNRAS.541.3280Y}. The alternatives include ring-shaped mountains on a hard crust, asymmetric quadrupolar mountains, and mountains on a pre-existing ocean, using an empirical equation of state for a degenerate electron gas \citep{1983ApJ...267..315P}, as well as a current-free outer boundary condition, which allows higher-order multipoles to develop and evolve. \citet{2025MNRAS.541.3280Y} also considered the effect of a changing Alfv\'en radius, as the magnetic dipole moment evolves during the burial process. They found that, for ring-shaped mounds, closed magnetic field contours (loops) do not form, because the spreading mound drags field lines north and south. In contrast, as a filled mound spreads, it drags field lines equatorward exclusively \citep{2004MNRAS.351..569P,2008MNRAS.386.1294V,2023MNRAS.526.2058R}. A mountain grown on top of a liquid ocean and atmosphere (e.g. \citet{2007ASSL..326.....H}) also sinks through the soft base, modifying the magnetic dipole moment and mass ellipticity.

In a filled or ring-shaped mountain built on a hard or soft surface, the mountain is supported from below against gravity by light, strongly magnetised material. Irrespective of the details, such a configuration is prone to MHD instabilities and subsequent cross-field transport, whether one uses \eqref{eq:dmdpsi_pm04} for $dM/d\psi$ or the arguably more realistic profiles specified by \citet{2025MNRAS.541.3280Y}. In this paper, we stick with \eqref{eq:dmdpsi_pm04} for the sake of simplicity, in order to keep the focus on the physical consequences of the semi-analytic mass transport scheme proposed by \citet{2020JPlPh..86f9002K}. A more general treatment is postponed to future work.

\subsection{Cross-field mass transport}\label{subsec: cfmf}
 In the semi-analytic picture proposed by \citet{2020JPlPh..86f9002K}, matter diffuses across magnetic flux surfaces in response to the nonlinear and nonideal development of the Schwarzschild instability. A spectrum of unstable modes is excited, forming a turbulent velocity field whose shear generates magnetic and density perturbations at smaller scales. At the resistive scale, flux freezing breaks down, and matter diffuses across magnetic flux surfaces in a manner that flattens the density gradient locally, i.e.\ nullifies \(d\rho/d\xi\).

In this paper, we implement mass transport between two flux tubes separated by an unstable flux surface surface \(\psi\) by moving an infinitesimal amount of mass \(\Delta M(\psi)\) from one flux tube to the next, such that the masses in the flux tubes \(\psi-\Delta \psi \leq \psi' \leq \psi\) and \(\psi \leq \psi' \leq \psi+\Delta\psi\) are equal after the mass moves. For brevity we refer to the region $\psi \leq \psi' \leq \psi+\Delta\psi$ as the flux tube $\psi$. When several unstable flux surfaces are adjacent to one another, forming an extended unstable region, we share the total mass in the region equally between its constituent flux tubes. The justification for the mass transport scheme, as well as a discussion of alternatives, is presented in \appendixautorefname~\ref{sec:app_A_mass_transport}. Specifically, it is shown in Appendix \ref{sec:app_A_mass_transport} that mass equalisation is equivalent exactly to density equalisation, when adjacent flux tubes are straight and have equal cross-sectional areas, and the two forms of equalisation remain approximately equivalent, when adjacent flux tubes are curved, over the relatively short scale height (compared to \(R_\ast\)) of a magnetic mountain.

As shown in Appendix \ref{sec:app_A_mass_transport}, for two adjacent flux tubes $\psi-\Delta\psi\leq \psi'\leq\psi$ and $\psi\leq\psi'\leq\psi+\Delta\psi$, the mass moved between them in order to equalise their masses is
\begin{equation}
    \Delta M(\psi) = \frac{1}{2}\left[\int_{\psi}^{\psi + \Delta\psi} d\psi'\left(\frac{dM}{d\psi'}\right)^{\mathrm{(init)}} - \int_{\psi - \Delta\psi}^{\psi}d\psi' \left(\frac{dM}{d\psi'}\right)^{\mathrm{(init)}} \right],
    \label{eq:delta_mass_transport_01}
\end{equation}
where a positive value of $\Delta M (\psi)$ corresponds to net mass transport from flux tube $\psi-\Delta\psi\leq \psi'\leq\psi$ to  $\psi\leq\psi'\leq\psi+\Delta\psi$, while a negative $\Delta M (\psi)$ corresponds to transport in the opposite direction. The mass-flux distribution may then be adjusted according to
\begin{equation}
    \left(\frac{dM}{d\psi}\right)^{\rm(adj)} = \left(\frac{dM}{d\psi}\right)^{\rm(init)} + \frac{\Delta M(\psi)}{\Delta\psi}.
    \label{eq:update_dmdpsi_accretion_delta_m_00}
\end{equation}
In Equations \eqref{eq:delta_mass_transport_01} and \eqref{eq:update_dmdpsi_accretion_delta_m_00}, the superscripts (init) (initial) and (adj) (adjusted) label the mass-flux distribution before and after cross-field mass transport respectively. 

\begin{figure}
    \includegraphics[width=\columnwidth]{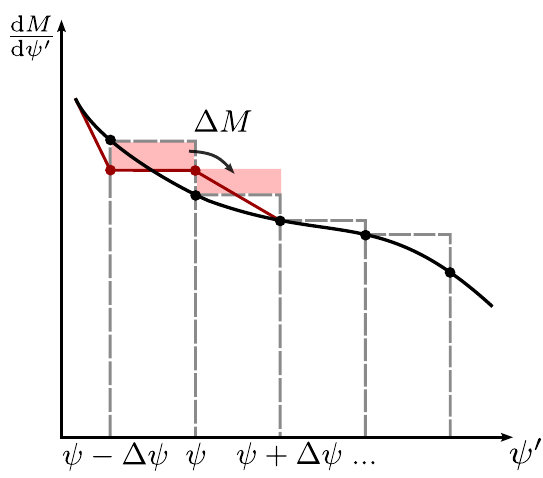}
    \caption{Cross-field mass transport: adjusting the initial mass-flux distribution (black curve) by moving mass \(\Delta M\) (red rectangle) across the unstable flux surface \(\psi' = \psi\), resulting in the adjusted distribution (red curve) given by \equationautorefname~\eqref{eq:update_dmdpsi_accretion_delta_m_00}. The curvature \(d^2M/d\psi'^2\) increases at the leading (\(\psi' = \psi + \Delta\psi\)) and trailing (\(\psi' = \psi - \Delta\psi\)) boundaries, resulting in the formation of new, smaller unstable regions, which are nullified in subsequent iterations.}
    \label{fig:illustration_01}
\end{figure}

Equation \eqref{eq:delta_mass_transport_01} states that the mass difference between two adjacent flux tubes, as in Figure \ref{fig:illustration_01}, is reapportioned equally between the tubes (e.g.\ subtract half from the left-hand tube, add half to the right-hand tube), to equalise the masses in the two tubes.
The adjustment is made at every flux surface, where the instability criterion in \equationautorefname~\eqref{eq:delta_criterion_01} is satisfied. The process is illustrated in \figureautorefname~\ref{fig:illustration_01} for mass transport across the (unstable) flux surface \(\psi' = \psi\). 
Once the mass-flux distribution is updated by applying \eqref{eq:delta_mass_transport_01} and \eqref{eq:update_dmdpsi_accretion_delta_m_00} to every unstable flux surface, Equation \eqref{eq:GS_eqn_03} is solved to find the new equilibrium, and the process repeats, until the instability disappears throughout the simulation volume. 

Should we expect a single iteration of mass transport to make all unstable points disappear? No. Applying \eqref{eq:delta_mass_transport_01} and \eqref{eq:update_dmdpsi_accretion_delta_m_00} nullifies the density gradients within the unstable region but not at the leading and trailing boundaries  ($\psi'=\psi+d\psi$ and $\psi'=\psi-d\psi$ respectively), where the curvature \(d^2M/d\psi^2\) increases, and new unstable regions develop, which are nullified in subsequent iterations of cross-field mass transport. Fracturing into multiple, smaller unstable regions is observed in the simulation output in Section \ref{sec:4}. We do not speculate whether or not fracturing occurs in reality, i.e.\ when the astrophysical system evolves according to time-dependent, nonideal MHD. In this paper, it occurs merely as an intermediate stage in the numerical implementation of the semi-analytic mass-transport scheme proposed by \citet{2020JPlPh..86f9002K}.
Once the instability disappears after enough iterations, and a new equilibrium  is computed at the end of one step in the quasistatic accretion sequence, further mass (distributed according to \eqref{eq:dmdpsi_pm04}) is accreted by adding $(dM/d\psi)^{\rm (acc)}$ to $(dM/d\psi)^{\rm (adj)}$ to obtain the updated mass-flux distribution
\begin{equation}
    \left(\frac{dM}{d\psi}\right)^{\rm (upd)} = \left(\frac{dM}{d\psi}\right)^{\rm (adj)} + \left(\frac{dM}{d\psi}\right)^{\mathrm{(acc)}}.
    \label{eq:update_dmdpsi_accretion_01}
\end{equation}
\subsection{Quasistatic approximation}\label{subsec:quasistatic_approx}
It is important to emphasise that the Grad-Shafranov equation \eqref{eq:GS_eqn_03} describes a steady-state magnetic mountain. Yet accretion onto the magnetic polar cap, which drives the evolution of a magnetic mountain --- including by triggering the Schwarzschild and other instabilities --- is a time-dependent process. In this paper, we study the astrophysically relevant regime $M_{\rm a} \gtrsim 10^{-5} M_\odot$, which corresponds to $\gtrsim 10^3 (\dot{M}_{\rm a} / 10^{-8} M_\odot \, {\rm yr^{-1}})^{-1} \, {\rm yr}$ of persistent accretion. A time-dependent MHD solver cannot handle the problem, due to the large dynamic range involved. How, therefore, should one apply the Grad-Shafranov equation \eqref{eq:GS_eqn_03} and flux-freezing constraint \eqref{eq:Fpsi_01} to approximate the time-dependent problem, where $M_{\rm a}$ increases to reach $\gtrsim 10^{-5} M_\odot$? In this respect, at least, the large dynamic range is helpful, as it justifies treating mountain evolution in terms of a quasistatic sequence of Grad-Shafranov equilibria. We describe why in the next paragraph.

The Schwarzschild instability is triggered, after relatively small amounts of matter are accreted. We find in Section \ref{sec:4} that unstable flux surfaces start to emerge for $M_{\rm a} \gtrsim 10^{-7} M_\odot$. \footnote{Early estimates that ballooning modes are triggered for $M_{\rm a} \gtrsim 10^{-12} M_\odot$ \citep{2001ApJ...553..788L} do not take into account the stabilising effects of the magnetic line tying boundary condition and Lorentz back reaction from the equatorial magnetic tutu via \eqref{eq:Fpsi_01}.} Once triggered, the instability develops nonlinearly on the local Alfv\'{e}n time-scale, $\tau_{\rm A} \sim 6.5~ (B/10^{12} \, {\rm G})^{-1} (\rho/ 10^{15} \, {\rm g\, cm^{-3}})^{1/2} \, {\rm s}$, which in turn is much shorter than the accretion time scale. Therefore it makes sense physically to proceed as follows: (i) add enough mass ($\sim 10^{-7} M_\odot$) to trigger the instability, by adjusting $dM/d\psi$ according to \eqref{eq:update_dmdpsi_accretion_01}; (ii) solve the Grad-Shafranov equation \eqref{eq:GS_eqn_03} and flux freezing constraint \eqref{eq:Fpsi_01} for the updated equilibrium, because the instability and associated cross-field mass transport occur almost instantaneously, on the time-scale $\tau_{\rm A} \ll M_{\rm a} / \dot{M}_{\rm a}$; (iii) add enough mass ($\sim 10^{-7} M_\odot$) to trigger the instability again, creating one or more unstable regions; and (iv) iterate until astrophysically relevant accreted masses $\gtrsim 10^{-5} M_\odot$ are reached. \footnote{Once magnetic bubbles form \citep{2004MNRAS.351..569P, 2007MNRAS.376..609P}, the mathematical and physical character of the Grad-Shafranov problem changes fundamentally \citep{1989ApJ...345.1034K}, and a new approach is needed, as noted by previous authors \citep{2008MNRAS.386.1294V,2013MNRAS.430.1976M,2013MNRAS.435..718M,2020JPlPh..86f9002K,2023MNRAS.526.2058R,2025MNRAS.541.3280Y}.} The more realistic alternative, which is to track the time-dependent evolution of both the instability and the accretion simultaneously with a three-dimensional resistive MHD solver, is prohibitively expensive computationally due to the large dynamic range, as noted above.

The reader may wonder whether the step-by-step recipe in the previous paragraph is mandatory. Can one do the calculation in ``one shot'' instead, taking advantage of the fact that the Grad-Shafranov equation describes a steady state? It turns out that the answer to this subtle physical question is no. Solutions to the Grad-Shafranov equation are not unique. They depend on the evolutionary path taken by $dM/d\psi$ and hence $F(\psi)$ on the way to reaching a particular value of $M_{\rm a}$. We test this explicitly in Section~\ref{sec:path_dependence}, where we calculate the instability-free equilibrium resulting from cross-field mass transport for $M_{\rm a} = 5\times10^{-7} M_\odot$ in two ways: (i) in one shot, by setting $dM/d\psi$ due to polar accretion according to \eqref{eq:dmdpsi_pm04} for $M_{\rm a}=5\times10^{-7} M_\odot$, calculating the unstable surfaces, and then nullifying the instabilities by one round of cross-field mass transport; and (ii) in ten increments of $5\times10^{-8} M_\odot$, to reach the same final $M_{\rm a}=5\times10^{-7} M_\odot$, but adjusting $dM/d\psi$ through polar accretion and instability-nullifying cross-field mass transport after every one of the ten increments. The results, presented in Section \ref{sec:path_dependence}, demonstrate unambiguously that path dependence produces different outcomes in cases (i) and (ii), although encouragingly the two outcomes are broadly similar.

\subsection{Numerical implementation}\label{sec:3_numerical_implementation}
We solve the Grad-Shafranov equation \eqref{eq:GS_eqn_03} and the flux-freezing condition \eqref{eq:Fpsi_01}, paired with the mass transport scheme \eqref{eq:delta_mass_transport_01}--\eqref{eq:update_dmdpsi_accretion_01}, using a modified version of the iterative numerical scheme developed by \citet{2004MNRAS.351..569P}. Details of the algorithm are contained in Appendix \ref{sec:app_B_numerical_implementation}. Here we summarise the main points for the convenience of the reader. We solve a dimensionless version of \eqref{eq:GS_eqn_03} on a grid of dimension \((N_r, N_\theta)\) with initial guesses for \(\psi(r,\theta)\) and \(dM/d\psi\), which fix the form of \(F(\psi)\) through \eqref{eq:Fpsi_01}. We generate solutions for $b = \psi_\ast/\psi_{\rm a}=3$ and 10, in keeping with \citet{2004MNRAS.351..569P}. We track \(N_c = N_r - 1\) flux surfaces, and solve equation \eqref{eq:GS_eqn_03} using successive over-relaxation. Details of the grid, the dimensionless variables, initial and boundary conditions, and the relaxation scheme are presented in Appendices \ref{sec:app_b1_grid_dim_vars}--\ref{sec:app_b3_GS_solver_SOR}. Once an equilibrium solution is found, unstable points on the \(N_c\) flux surfaces are identified using \eqref{eq:delta_criterion_00}, and the mass transport scheme adjusts the form of \(dM/d\psi\) via \eqref{eq:delta_mass_transport_01} and \eqref{eq:update_dmdpsi_accretion_delta_m_00}. Once zero unstable points remain, and the instability is nullified, more mass is accreted, until the instability is triggered again. Convergence of the solution is discussed in Appendix \ref{sec:app_b7_convergence}. The calculation of the instability criterion \eqref{eq:delta_criterion_00} is presented in Appendix \ref{sec:app_b4_instab_calc}, whilst dimensionless forms of \eqref{eq:delta_mass_transport_01}--\eqref{eq:update_dmdpsi_accretion_01} are presented in Appendices \ref{sec:app_b5_cfmf_calc}--\ref{sec:app_b6_mass_accretion}. 
%

\section{Nullifying the instability}{\label{sec:4}}
In this section, we present results from executing the numerical scheme in Section \ref{sec:3} and Appendix \ref{sec:app_B_numerical_implementation} for a range of $M_{\rm a}$ values. The results fall into three categories: (i) identifying unstable regions through \eqref{eq:delta_criterion_00} and \eqref{eq:delta_crit_01} as a function of $M_{\rm a}$ (Section \ref{sec:unstable_region}); (ii) nullifying the instability through cross-field mass transport by iterating \eqref{eq:delta_mass_transport_01}--\eqref{eq:update_dmdpsi_accretion_01} (Section \ref{sec:cf_trans_dissipates_region}) and investigating the path dependence inherent in the quasistatic approximation (Section \ref{sec:path_dependence}); and (iii) calculating the final hydromagnetic equilibrium, after the instability is nullified, as a function of $M_{\rm a}$ up to the representative value $M_{\rm a} = 10^{-5}M_\odot$ before magnetic bubbles form (Section \ref{sec:hydromag_structure_of_mtn}). The results in this section establish a foundation for calculating astrophysical observables such as the magnetic dipole moment and mass quadrupole moment as functions of $M_{\rm a}$ in Section \ref{sec:astro_observables}. In what follows, we adopt the following fiducial values to facilitate comparison with \citet{2004MNRAS.351..569P}:
$M_\ast = 1.4M_\odot,\,$\(R_\ast = 10^4~\rm m\), \(B_\ast = 10^8~\rm T\), and \(c_s = 10^6 ~\rm m~s^{-1}\).
\subsection{Unstable region}\label{sec:unstable_region}

We start by calculating the unstable region for initial masses in the range $10^{-8} \leq M_{\rm a}^{(\rm init)} / M_\odot \leq 10^{-5}$. That is, we calculate the hydromagnetic equilibrium satisfying \eqref{eq:GS_eqn_03} and \eqref{eq:Fpsi_01} in one shot, with $dM/d\psi$ given by \eqref{eq:dmdpsi_pm04} for the selected initial mass $M_{\rm a}^{(\rm init)}$, and then identify the unstable points within the equilibrium satisfying \eqref{eq:delta_criterion_00} and \eqref{eq:delta_crit_01}. For the purpose of this one-shot exercise, whose goal is to illustrate how the numerical recipe in Section \ref{sec:3} and Appendix \ref{sec:app_B_numerical_implementation} works in practice, we do not build up the accreted mass quasistatically (see Section \ref{subsec:quasistatic_approx}). Quasistatic build up is analysed in Sections \ref{sec:path_dependence} and \ref{sec:astro_observables}.

\begin{figure*}
\centering
\includegraphics[width=\linewidth]{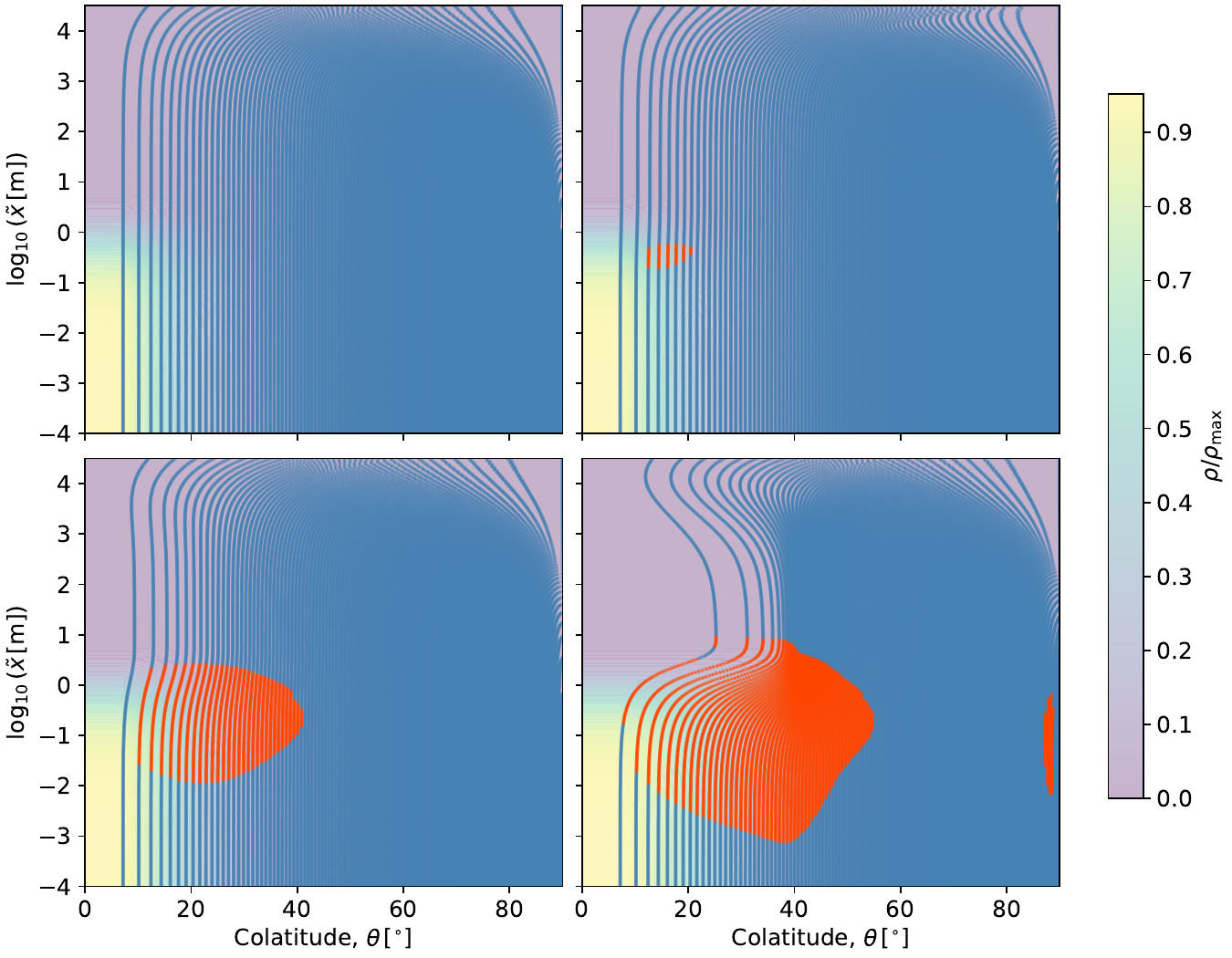}
\caption{Unstable region as a function of the initial accreted mass, with $M_{\rm a}^{\rm (init)}=10^{-8}M_\odot$, $\rho_{\rm max}=5.0\times10^{14}\,\mathrm{kg\,m^{-3}}$ (top left); $M_{\rm a}^{\rm (init)}=10^{-7}M_\odot$, $\rho_{\rm max}=4.9\times10^{15}\,\mathrm{kg\,m^{-3}}$ (top right); $M_{\rm a}^{\rm (init)}=10^{-6}M_\odot$, $\rho_{\rm max}=4.0\times10^{16}\,\mathrm{kg\,m^{-3}}$ (bottom left); and $M_{\rm a}^{\rm (init)}=10^{-5}M_\odot$, $\rho_{\rm max}=1.9\times10^{17}\,\mathrm{kg\,m^{-3}}$ (bottom right). Stability is determined via \eqref{eq:delta_criterion_00}. In each panel, unstable and stable points are coloured orange and blue respectively; points trace magnetic field lines. The yellow background shading traces density contours, indicating the position of the mountain. The density contours of each panel are normalised by their respective values of $\rho_{\rm max}$, and the resulting normalised contour values are indicated by the colour bar.}
\label{fig:unstable_region_with_densities}
\end{figure*}

Figure \ref{fig:unstable_region_with_densities} displays the unstable region in four panels corresponding to $M_{\rm a}^{({\rm init})} / M_\odot = 10^{-8}$, $10^{-7}$, $10^{-6}$, and $10^{-5}$. In each panel, unstable and stable points are coloured orange and blue respectively. The points trace magnetic field lines; the yellow background shading traces density contours. There are zero unstable points in the top-left panel, corresponding to $M_{\rm a}^{(\rm init)} / M_\odot = 10^{-8}$; we find that the instability is triggered for $M_{\rm a}^{(\rm init)} / M_\odot \geq 9 \times 10^{-8}$. In the top right panel, corresponding to $M_{\rm a}^{({\rm init})} / M_\odot = 10^{-7}$, the unstable region is relatively small and occurs at the leading edge of the accreted mountain, within approximately one scale height ($\sim x_0 = c_s^2R_\ast^2/GM_\ast$) of the surface; note the logarithmic vertical scale. The instability is triggered, even though the magnetic field lines are approximately straight within the bulk of the mountain. In the bottom two panels, the unstable region is larger. It is still centred on the leading edge of the accreted mountain but extends over tens of degrees of colatitude, e.g.\ $5^\circ \lesssim \theta \lesssim 55^\circ$ for $M_{\rm a}^{({\rm init})} / M_\odot = 10^{-5}$, and rises to approximately 10 scale heights ($\sim 10x_0$).

\begin{figure}
\includegraphics[width=\columnwidth]{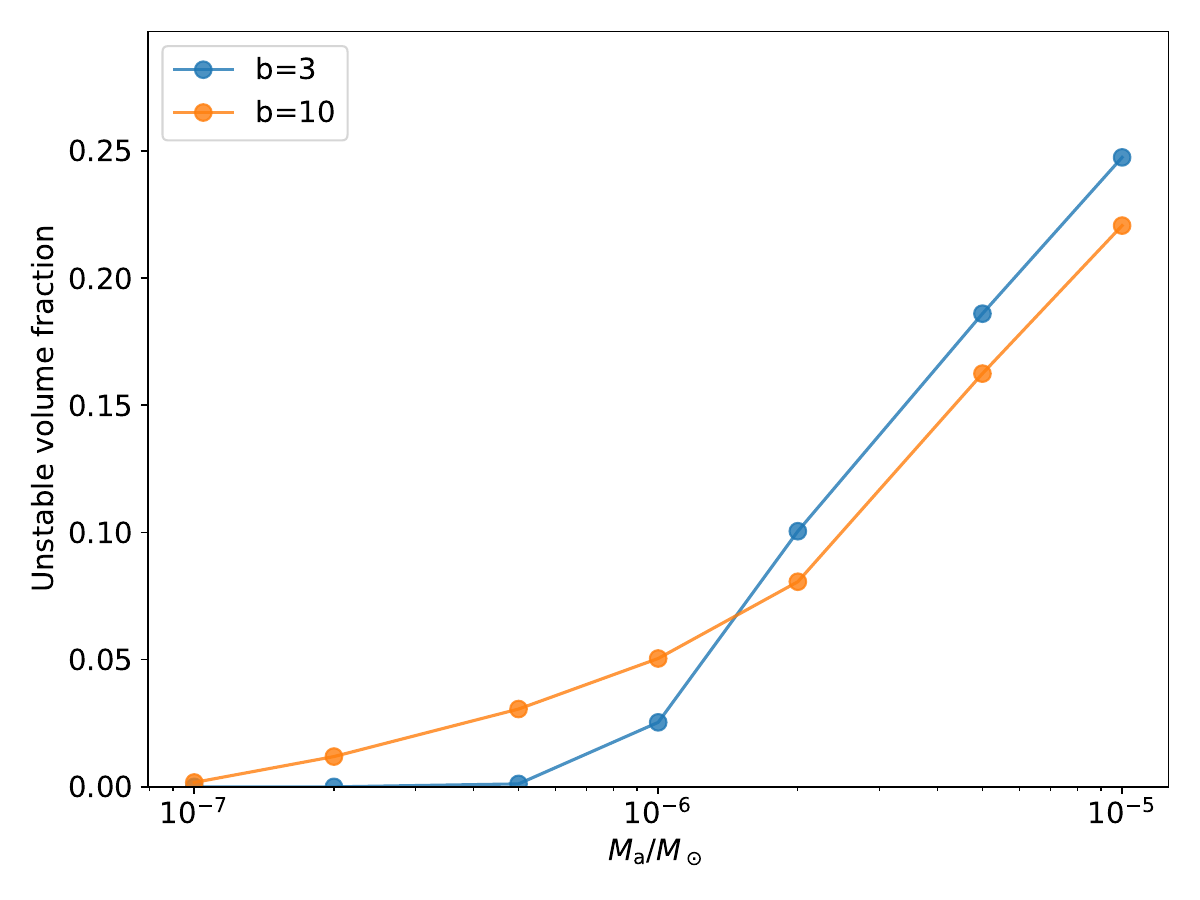}
\caption{Unstable fraction of the total flux tube volume within ten scale heights of the stellar surface, as a function of the initial accreted mass \(M_{\rm a}^{\rm (init)}\). Mass is distributed according to the mass flux distribution \eqref{eq:dmdpsi_pm04}, for accreted masses \(10^{-7}M_\odot\leq M_{\rm a} \leq 10^{-5}M_\odot\). The two curves are for \(b=3\) (blue) and \(b=10\) (orange). In these one-shot experiments (see Section \ref{sec:unstable_region}), the unstable region grows monotonically with $M_{\rm a}^{({\rm init})}$.}
\label{fig:unstable_vol_pctg}
\end{figure}

We quantify the extent of the unstable region by calculating the fraction of the flux tube volume $0\leq\psi\leq\psi_\ast$ below ten scale heights that is unstable, as a function of $M_{\rm a}^{({\rm init})}$. The results are displayed in Figure~\ref{fig:unstable_vol_pctg}. As expected, the unstable volume increases monotonically with $M_{\rm a}^{({\rm init})}$. Moreover, the unstable volume fraction is larger for $b=3$ (wider polar cap) than $b=10$, but the absolute difference is modest, viz.\ $\leq 0.002 $ for $10^{-7} \leq M_{\rm a}^{({\rm init})} / M_\odot \leq 10^{-5}$. Importantly, the unstable volume is much smaller than the closed field line region and smaller than but comparable to the volume of the accreted mountain even in the unrealistic one-shot scenario, where $M_{\rm a}^{({\rm init})}$ is loaded instantaneously. This is one factor that explains why previous studies with time-dependent MHD simulations have found that magnetic mountains are not disrupted completely, even when their two-dimensional Grad-Shafranov equilibria are formally unstable \citep{2007MNRAS.376..609P,2008MNRAS.386.1294V, 2009MNRAS.395.1985V,2013MNRAS.430.1976M,2013MNRAS.435..718M}. This conclusion strengthens, when the mountain accretes quasistatically, as confirmed by the multi-shot simulations in Sections \ref{sec:hydromag_structure_of_mtn} and \ref{sec:astro_observables}.

Relatedly, we confirm the finding of \citet{2004MNRAS.351..569P}, that single-shot equilibria satisfying \eqref{eq:GS_eqn_03}, \eqref{eq:Fpsi_01} and \eqref{eq:dmdpsi_pm04} do not converge for  $M_{\rm a}^{({\rm init})} / M_\odot \gtrsim 3\times 10^{-5}$, because flux surfaces intersect at X-points, and magnetic bubbles form; see \citet{1989ApJ...345.1034K}, Section 4.7 in \citet{2004MNRAS.351..569P}, and Section 4.1 in \citet{2008MNRAS.386.1294V}.

\subsection{Iterative cross-field mass transport}\label{sec:cf_trans_dissipates_region}
\begin{figure*}
    \centering
    \includegraphics[width=\linewidth]{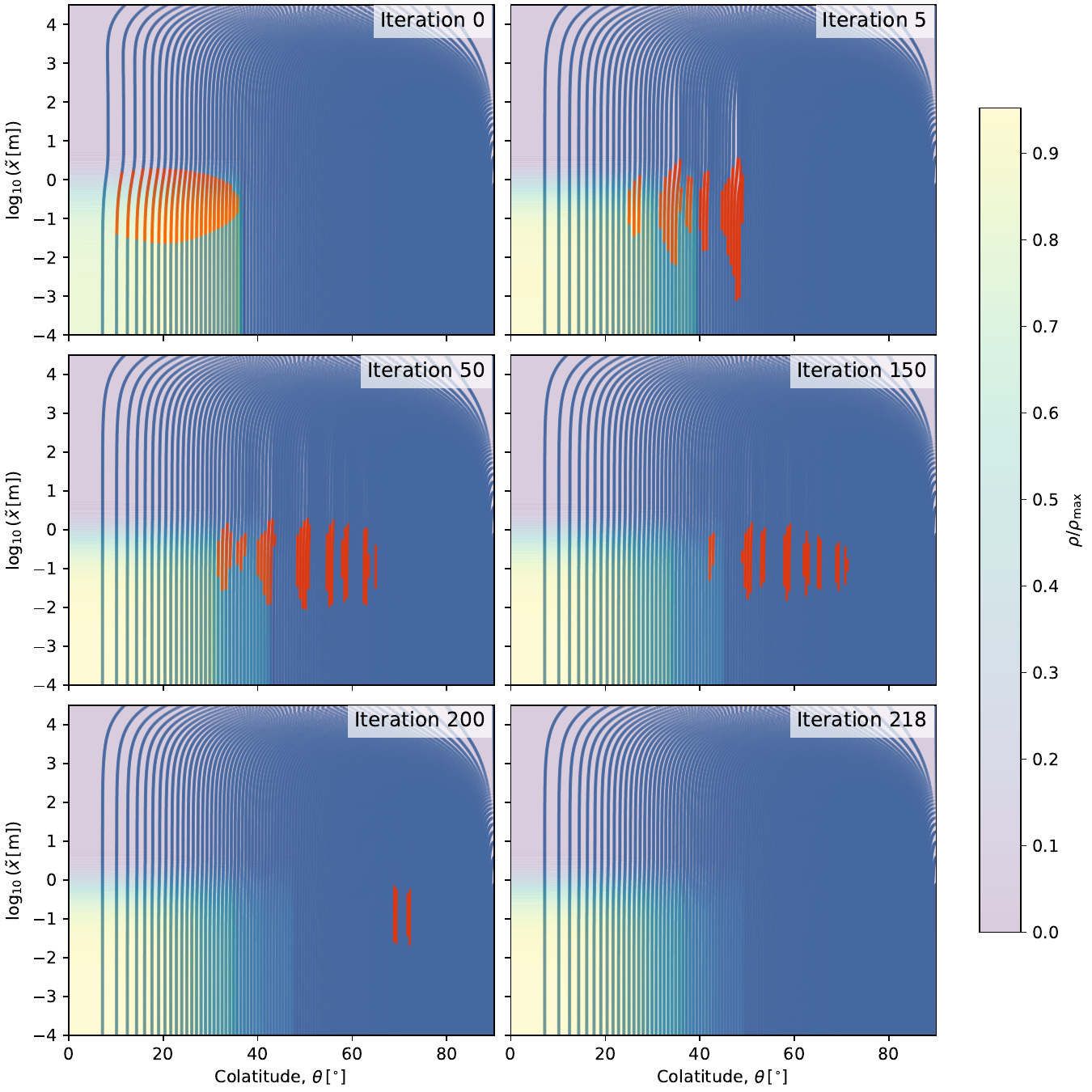}
    \caption{Nullifying the instability through cross-field mass transport: iteration of an unstable region for $M_{\rm a} = 5\times10^{-7}M_\odot$, $b=10$, according to the numerical scheme proposed by \citet{2020JPlPh..86f9002K} and described in Section \ref{sec:3} and Appendix \ref{sec:app_B_numerical_implementation}. The iteration number corresponding to each snapshot is quoted in the legend at the top right of each panel. The instability is nullified after 218 iterations. The format of each panel is the same as in Figure \ref{fig:unstable_region_with_densities}, viz.\ orange and blue points are unstable and stable respectively. Here one finds \(\rho_{\rm max}=6.64\times 10^{15}\mathrm{kg\,m^3}\).}
    \label{fig:ma_xx_dissipate_unstable_region_subplots}
\end{figure*}

We now demonstrate that the cross-field mass transport scheme proposed by \citet{2020JPlPh..86f9002K} and described in Section \ref{subsec: cfmf} succeeds in nullifying the instability wherever it occurs and restores the unstable region in Figure \ref{fig:unstable_region_with_densities} to a state of marginal stability. We find that the nullification occurs iteratively in the same way throughout the range $10^{-8} \leq M_{\rm a}^{({\rm init})} / M_\odot \leq 10^{-5}$, so we present results for one particular, representative case, viz. \ $M_{\rm a}^{({\rm init})} / M_\odot = 5\times10^{-7}$. The starting point is a one-shot equilibrium, depicted in the top left panel of Figure \ref{fig:ma_xx_dissipate_unstable_region_subplots}, which is adequate for demonstrating the operation of the numerical scheme. Astrophysically realistic multi-shot equilibria, where the mountain accumulates quasistatically, are studied in Sections \ref{sec:hydromag_structure_of_mtn} and \ref{sec:astro_observables}.

Figure \ref{fig:ma_xx_dissipate_unstable_region_subplots} displays eight snapshots drawn from 218 iterations of the numerical scheme for $M_{\rm a}^{({\rm init})} / M_\odot = 5\times10^{-7}$. Upon comparing the early snapshots with the starting state, we observe that the unstable region breaks up under iteration into multiple, disconnected subregions, because cross-field mass transport nullifies the instability on some flux surfaces at the expense of exacerbating mass imbalances on other flux surfaces. Loosely speaking, this is tantamount to flattening a wrinkle in a rug; flattening it in one place causes it to ``pop up'' somewhere else. We have no way of knowing at present, whether this behaviour also occurs in the true, time-dependent MHD evolution; the panels in Figure \ref{fig:ma_xx_dissipate_unstable_region_subplots} depict iterations in a numerical scheme, not snapshots in time during the true, physical evolution. The multiple unstable regions move equatorwards, simultaneously shrinking. After approximately 100 iterations they reach equatorial latitudes and either disappear or merge into a single unstable region, after 200 iterations, which disappears at iteration 218, whereupon the mountain is marginally stable everywhere. Interestingly, the density profile and magnetic field structure of the marginally stable mountain are similar to the starting state. That is, stability is restored with modest adjustments to the hydromagnetic structure, consistent with previous work \citep{2007MNRAS.376..609P,2008MNRAS.386.1294V,2009MNRAS.395.1985V,2013MNRAS.435..718M}.

\begin{figure}
\includegraphics[width=\columnwidth]{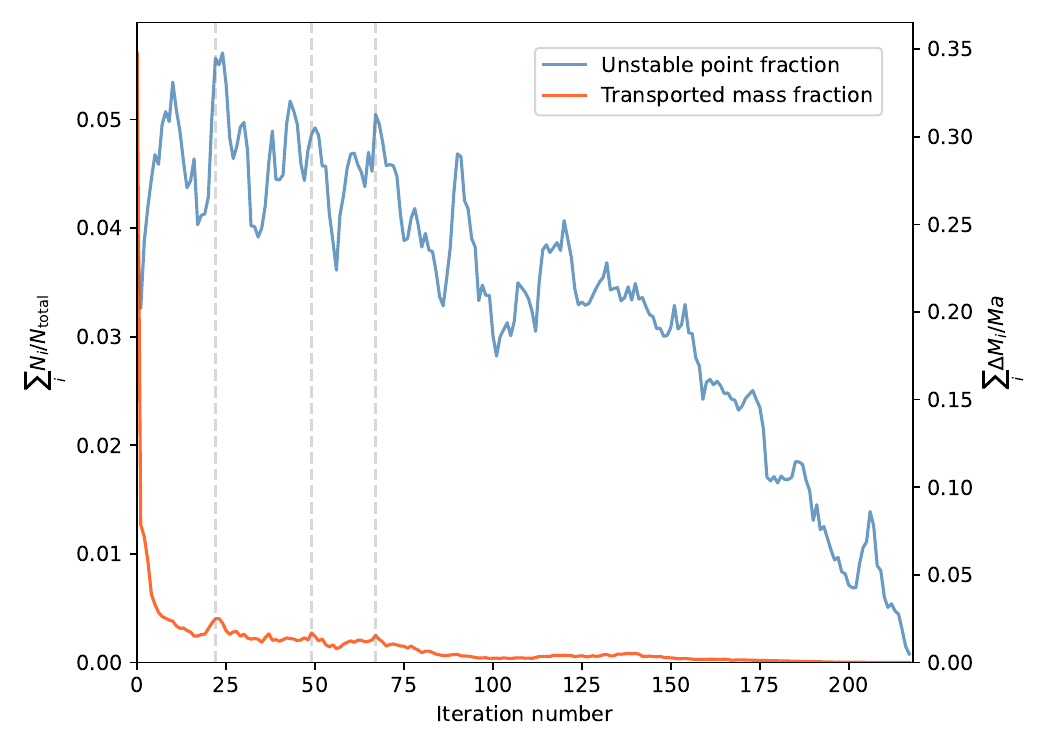}
\caption{Iterative progress towards nullifying the instability: fraction of points \(\sum_i N_i/N_{\rm total}\) that are unstable, as a function of iteration (left axis); and the fraction of the total accreted mass \(\sum_i \Delta M_i /M_{\rm a}\) that is transported, as a function of iteration (right axis). The subscript $i$ labels flux surfaces: $N_i$ is the number of unstable points on flux surface $\psi_i$, and $N_{\rm total}$ is the total number of points on all flux surfaces, stable or unstable. We test for instability using \eqref{eq:delta_criterion_00} and \eqref{eq:delta_crit_01}. Similarly, $\Delta M_i$ is the mass transported across flux surface $\psi_i$. Physical parameters: \(M_{\rm a}=5\times10^{-7}M_\odot\), $b=10$. Dashed lines correspond to iterates, where multiple smaller unstable regions temporarily coalesce into larger, less numerous regions, decreasing $N_i$ and increasing $\Delta M_i$ temporarily. Ultimately, the unstable region disappears after 218 iterations.}
\label{fig:unstable_region_ma_1e-7_plot_01}
\end{figure}

 To quantify the degree of iterative adjustment further, we plot in Figure \ref{fig:unstable_region_ma_1e-7_plot_01} the total number of unstable points as a function of iteration number, together with the mass transported across magnetic field lines at each iteration normalised by $M_{\rm a}^{({\rm init})}$. We see that the number of unstable points rises during the early iterations, as the unstable region breaks up into multiple, disconnected, unstable subregions, which move equatorwards collectively and grow in volume overall. Simultaneously, a significant fraction ($\sim35$ per cent) of the mountain mass undergoes transport, though this quickly reduces to less than $5$ per cent after $\sim10$ iterations. After $\sim 10$ iterations, the trend reverses. The number of unstable points decreases steadily, punctuated by temporary setbacks, when adjacent unstable regions temporarily coalesce (creating a larger, unified region before fracturing again into multiple smaller regions by subsequent mass transport). Simultaneously, the total transported mass continues to decline, punctuated by small jumps when the regions temporarily coalesce. This can be seen in Figure \ref{fig:unstable_region_ma_1e-7_plot_01}, where the small bumps in the transported mass fraction (red curve) align with the peaks of the unstable point fraction (blue) at iterations $\sim 22, 50$, and $68$ (indicated by the dashed grey lines). At these iterations, the temporarily larger regions boost the mass transport, driving the number of unstable points lower than before the coalescence. Overall, even at their peaks, the number of unstable points and the mass transported per iteration amount to small perturbations of the whole mountain. That is, the mountain adjusts locally; it is not disrupted in its entirety. This behaviour holds throughout the range $10^{-8} \leq M_{\rm a}^{({\rm init})} / M_\odot \leq 10^{-5}$.

\subsection{Path dependence}\label{sec:path_dependence}
The quasistatic recipe for accreting from scratch a mountain with final mass $M_{\rm a}$, described in Section \ref{subsec:quasistatic_approx}, is an approximation to the true, time-dependent MHD evolution in an astrophysical setting, which cannot be simulated in practice due to the extreme time-scale separation. Hence it is interesting to check whether the final, post-accretion equilibrium state, as calculated numerically, depends sensitively on the path taken to reach the final state. To that end, in this section, we compare $dM/d\psi$ for two representative equilibria with $M_{\rm a} = 5\times10^{-7} M_\odot$. One equilibrium is calculated in one shot, by solving \eqref{eq:GS_eqn_03}, \eqref{eq:Fpsi_01}, and \eqref{eq:dmdpsi_pm04} simultaneously for $M_{\rm a} = M_{\rm a}^{({\rm init})} = 5\times 10^{-7} M_\odot$ and then executing cross-field mass transport to nullify the unstable region, as in Sections \ref{sec:unstable_region} and \ref{sec:cf_trans_dissipates_region}. The hydromagnetic structure before nullifying the instability is depicted in the top right panel of Figure \ref{fig:unstable_region_with_densities}. The second equilibrium is calculated quasistatically in multiple steps. We start from $M_{\rm a}^{({\rm init})} = 5\times10^{-8} M_\odot$, solve \eqref{eq:GS_eqn_03}, \eqref{eq:Fpsi_01}, and \eqref{eq:dmdpsi_pm04} simultaneously, execute cross-mass field mass transport to nullify the unstable region, and obtain an intermediate estimate of $dM/d\psi$. Once the instability is nullified, we accrete an additional $5\times10^{-8} M_\odot$ atop the intermediate $dM/d\psi$, solve \eqref{eq:GS_eqn_03}, \eqref{eq:Fpsi_01}, and \eqref{eq:dmdpsi_pm04} simultaneously, execute cross-field mass transport, and update the intermediate estimate of $dM/d\psi$. We iterate the procedure 10 times, accreting $5\times10^{-8} M_\odot$ at each iteration, until we accumulate $M_{\rm a} = 5\times10^{-7} M_\odot$. 

\begin{figure}
    \centering
    \includegraphics[width=\columnwidth]{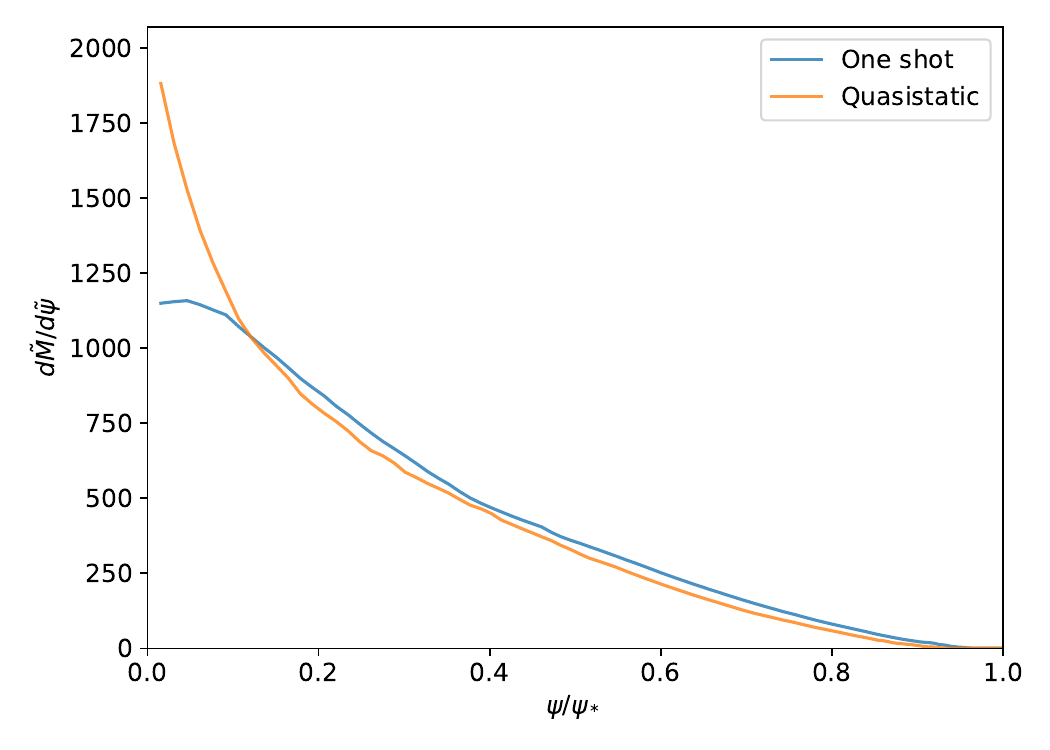}
    \caption{Path dependence of the mass-flux distribution: $dM/d\psi$ for two marginally stable mountains with $M_{\rm a} = 5\times10^{-7} M_\odot$ after cross-field mass transport. One mountain (blue curve) is built up in one shot, with $M_{\rm a}^{({\rm init})} = 5\times10^{-7} M_\odot$. The other mountain (orange curve) is built up quasistatically, by accreting a total of $M_{\rm a} =5\times10^{-7} M_\odot$ in ten equal increments, each $5\times10^{-8} M_\odot$, and nullifying the unstable region by cross-field mass transport after every increment. Axis label notation: $dM/d\psi = M_{\rm a}/\psi_0 (d\tilde{M}/d\tilde{\psi})$ (see Appendix \ref{sec:app_b1_grid_dim_vars} for details).}
    \label{fig:path-dependence}
\end{figure}

The mass-flux distributions for the one-shot and quasistatic mountains are plotted together in Figure~\ref{fig:path-dependence}. They are broadly similar, except for in the polar region, where they differ by $\sim38$ per cent at most at $\psi/\psi_\ast=0$. Here, we observe a polar plateau in the one-shot $dM/d\psi$, which is absent from the quasistatic $dM/d\psi$. This implies that the instability is nullified more economically, with less material transported equatorwards across magnetic field lines, when the quasistatic approximation is applied. We emphasise again that there is no way to know whether the quasistatic calculation matches the true, physical evolution without conducting time-dependent MHD simulations, which are prohibitive computationally at the time of writing. We observe the polar plateau in the one-shot $dM/d\psi$ consistently throughout the range $10^{-8} \leq M_{\rm a} / M_\odot \leq 10^{-5}$. 

\subsection{Hydromagnetic structure of a representative quasistatic mountain with \texorpdfstring{$M_{\rm a} = 10^{-5} M_\odot$}{Ma = 1e-5 Msun}}\label{sec:hydromag_structure_of_mtn}
We now present the hydromagnetic structure of a quasistatically assembled mountain stabilised by cross-field mass transport, and compare the results to studies published previously. We first calculate $dM/d\psi$ at several intermediate stages between an initial accreted mass $M_{\rm a}^{\rm(init)} = 10^{-8}M_\odot$ prior to the onset of hydromagnetic instabilities, and the final, representative mass $M_{\rm a} = 10^{-5}M_\odot$. We find that the stabilised intermediate $dM/d\psi$ profiles are approximated accurately by a universal function of $M_{\rm a}$ and $b$, allowing us to generate profiles at even larger accreted masses without resorting to the increasingly expensive iterative process of quasistatic assembly used during the calibration stage.

\begin{figure}
\centering
\includegraphics[width=\columnwidth]{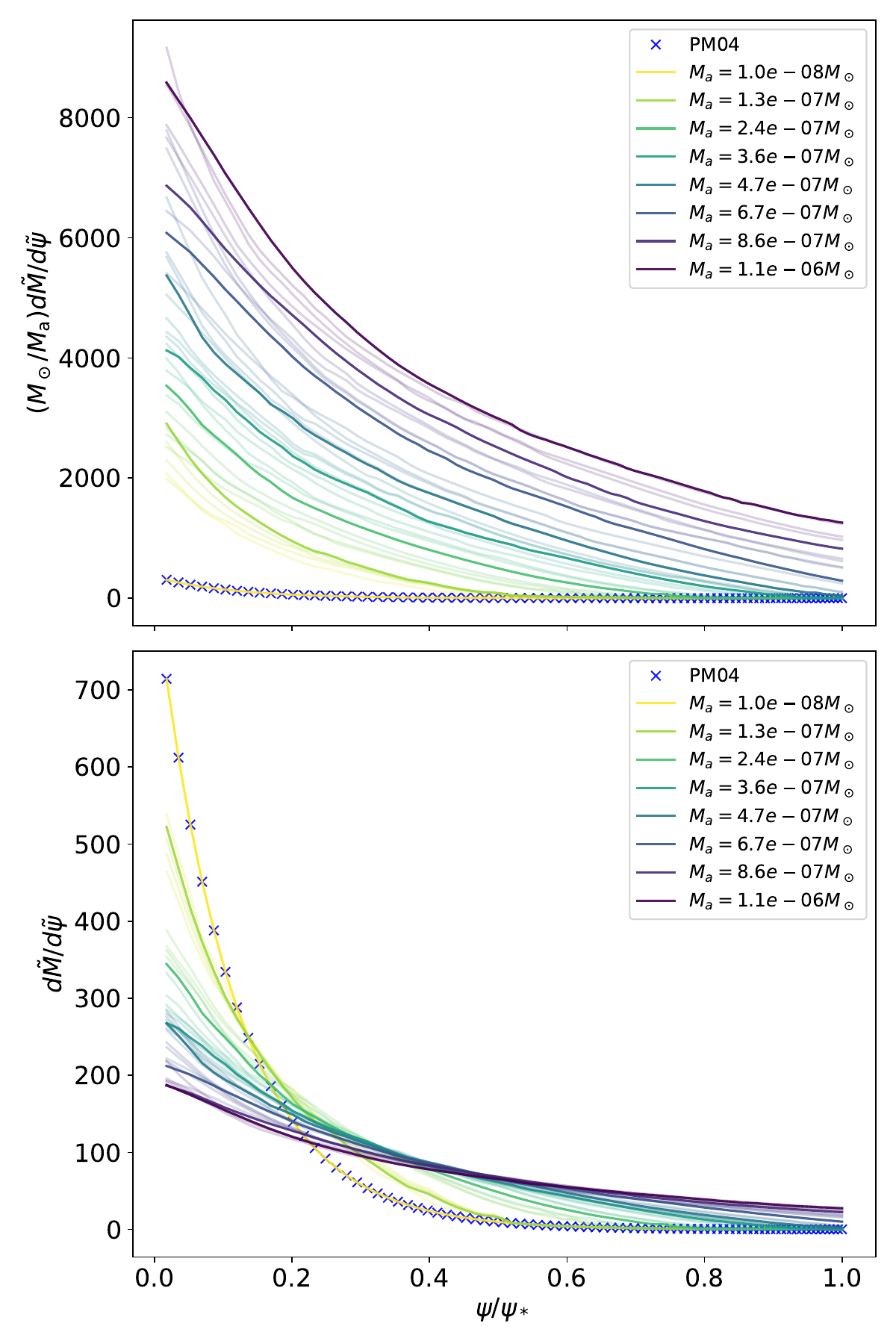}
\caption{Quasistatic assembly of the mountain: mass-flux distribution $d\tilde{M}/d\tilde{\psi}$ at intermediate iterations, after mass transport stabilises the mountain for that iteration and before accreting further mass at the next iteration. The top panel shows the mass-flux distributions prior to normalisation by the total accreted mass, and includes the initial PM04 profile (blue crosses) corresponding to Equation~\eqref{eq:dmdpsi_pm04} with $M_{\rm a} = 10^{-8}M_\odot$. The bottom panel replots the same distributions normalised to the corresponding accreted mass $M_{\rm a}'$. The normalised profiles are approximated empirically by Equation \eqref{eq:parameter_fit_sech}. Axis label notation: $dM/d\psi = M_{\rm a}/\psi_0 (d\tilde{M}/d\tilde{\psi})$ (see Appendix \ref{sec:app_b1_grid_dim_vars} for details).}
\label{fig:dmdpsi_evolution_b10_01}
\end{figure}

The top panel of \figureautorefname~\ref{fig:dmdpsi_evolution_b10_01} displays \(dM/d\psi\) as a function of $\psi$ from \(M_{\rm a}^{\rm (init)} = 10^{-8}M_\odot\) to \(M_{\rm a} = 1.1\times10^{-6}M_\odot\), at six intermediate masses $10^{-8} \leq M_{\rm a}' / M_\odot \leq 1.1\times10^{-6}$ (see legend), for $b=10$. The six profiles are plotted after cross-field mass transport stabilises the mountain. We terminate the quasistatic assembly at $M_{\rm a} = 1.1\times10^{-6}M_\odot$ due to the computational cost. We find that $dM/d\psi$ increases by a factor of 25 at $\psi = 0$ and 52 at $\psi = \psi_{\rm a}$, as the accreted mass $M_{\rm a}'$ increases from \(10^{-8}M_\odot\) to \(10^{-6}M_\odot\). In the bottom panel of ~\figureautorefname~\ref{fig:dmdpsi_evolution_b10_01}, we normalise each curve by $M_{\rm a}'/M_\odot$ and find that the normalised profile is approximated empirically by
\begin{equation}
    \left(\frac{d\tilde{M}}{d\tilde{\psi}}\right)_{\mathrm{(norm)}} = \frac{K_0}{2\psi_{\rm a} f(M_{\rm a}')}
    \frac{e^{-\psi/[\psi_{\rm a}f(M_{\rm a}')]}}{1 - e^{-\psi_\ast/[\psi_{\rm a}f(M_{\rm a}')]}} ,
    \label{eq:parameter_fit_sech}
\end{equation}
 with $K_0 = 2.1$,
 \begin{equation}
    f(M_{\rm a}) = 
    \begin{cases}
    1 & M_{\rm a}' < M_{\rm c, inst}(b)\\
    1 + K_1 b^{p}[(M_{\rm a}' - M_{\rm c, inst})/M_\odot]^{q} & M_{\rm a}' \geq M_{\rm c, inst}(b),
    \end{cases}
    \label{eq:parameter_fit_f_func}
 \end{equation}
$K_1 = 478$, $p = 1.73$, and $q = 0.65$, where $M_{\rm c, inst}$ is the critical mass at which the instability first appears, which is a function of $b$. The piecewise form of $f(M_{\rm a}')$ in \eqref{eq:parameter_fit_f_func} expresses the fact, that $dM/d\psi$ is unchanged from \eqref{eq:dmdpsi_pm04}, until enough mass accretes to trigger the instability and cross-field mass transport. For $b=10$ (corresponding to the $dM/d\psi$ curves in both panels of Figure \ref{fig:dmdpsi_evolution_b10_01}), we find $M_{\rm c, inst}/M_\odot = 9\times10^{-8}$, whereas for a wider accretion column (e.g. $b=3$) the instability appears at a higher mass, $M_{\rm c, inst}/M_\odot = 5.5\times10^{-7}$. The detailed structure of the intermediate $dM/d\psi$ distributions is examined for completeness in Appendix \ref{sec:app_C_dmdpsi_fit}. Equations \eqref{eq:parameter_fit_sech} and \eqref{eq:parameter_fit_f_func} are important in practical terms. Once calibrated, they let one generate stable mountain profiles without explicitly performing cross-field mass transport iteratively; the ultimate result of cross-field mass transport is captured approximately by \eqref{eq:parameter_fit_sech} and \eqref{eq:parameter_fit_f_func}. 

\begin{figure}
\centering
\includegraphics[width=\columnwidth]{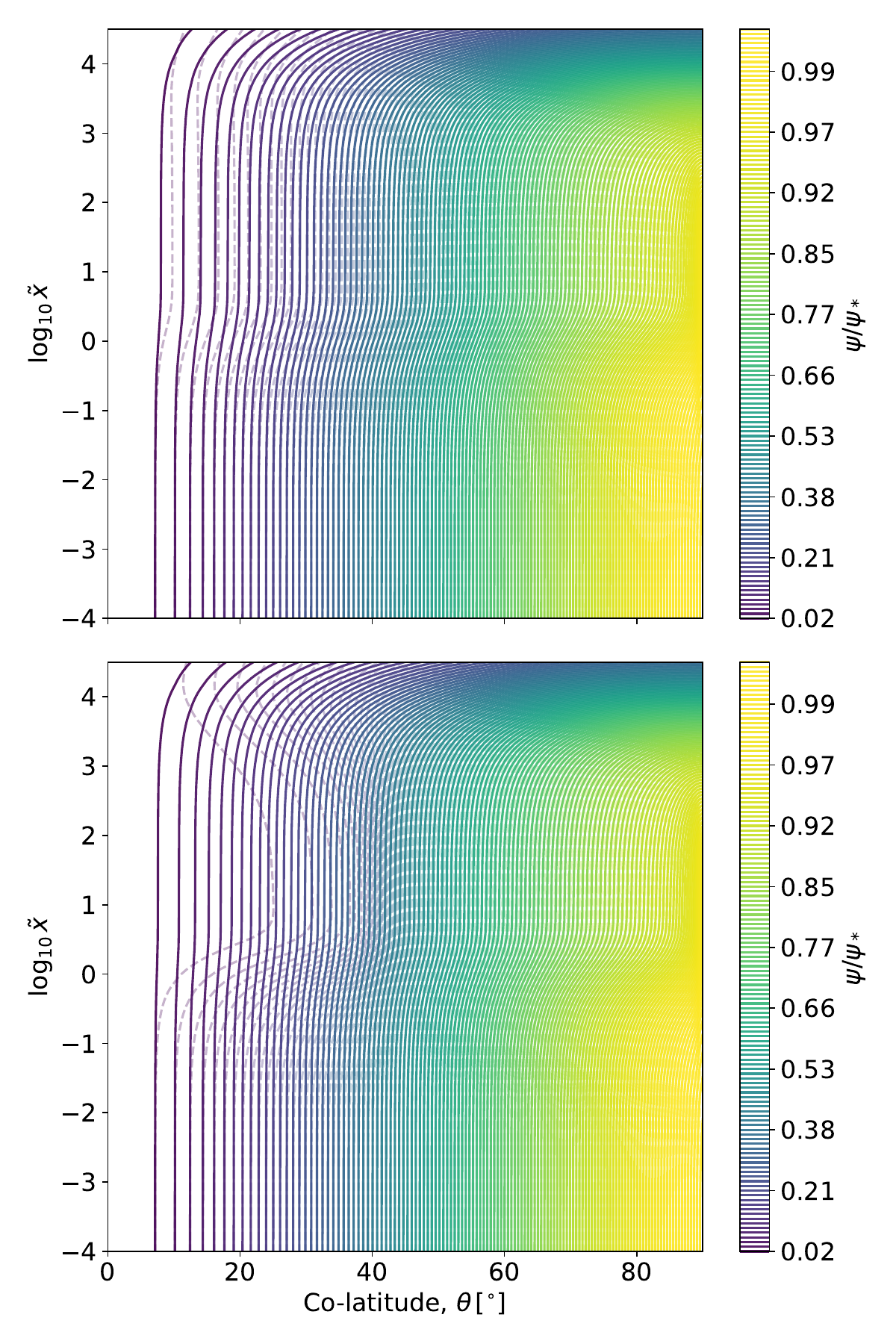}
\caption{Magnetic flux surfaces of a representative mountain with $M_{\rm a} = 1\times 10^{-5} M_\odot$ before (dashed curves) and after (solid curves) cross-field mass transport. The top panel is for $b=3$, the bottom for $b=10$. The contour colour corresponds to the value $\psi/\psi_\ast$.}
\label{fig:psi_ma_1e-5}
\end{figure}

We now apply \eqref{eq:parameter_fit_sech} and \eqref{eq:parameter_fit_f_func} to study the hydromagnetic structure of a representative mountain with $M_{\rm a}/M_\odot = 10^{-5}$ and compare the result with the corresponding mountain without mass transport in \citet{2004MNRAS.351..569P}. Figure~\ref{fig:psi_ma_1e-5} plots the magnetic flux surfaces in cross-section before (dashed) and after (solid) cross-field mass transport, for $b=3$ (top panel) and $b=10$ (bottom panel). We see that the magnetic flux surfaces shift towards the pole, partly relaxing the equatorial magnetic tutu observed by \citet{2004MNRAS.351..569P}. However, the relaxation is incomplete; the equilibrium does not return all the way to the undistorted, pre-accretion dipole. Before cross-field mass transport, the flux surfaces are significantly distorted in a region confined to co-latitudes $\theta\lesssim 50^{\circ}$. After cross-field mass transport, flux surfaces all the way to the equator are distorted, and screening currents $\mu_0^{-1}\nabla\times\mathbf{B}$ extend to the equator. However, the screening currents are weaker, and the magnetic dipole moment does not diminish by as much as in \cite{2004MNRAS.351..569P}.

\begin{figure*}
\centering
\includegraphics[width=\textwidth]{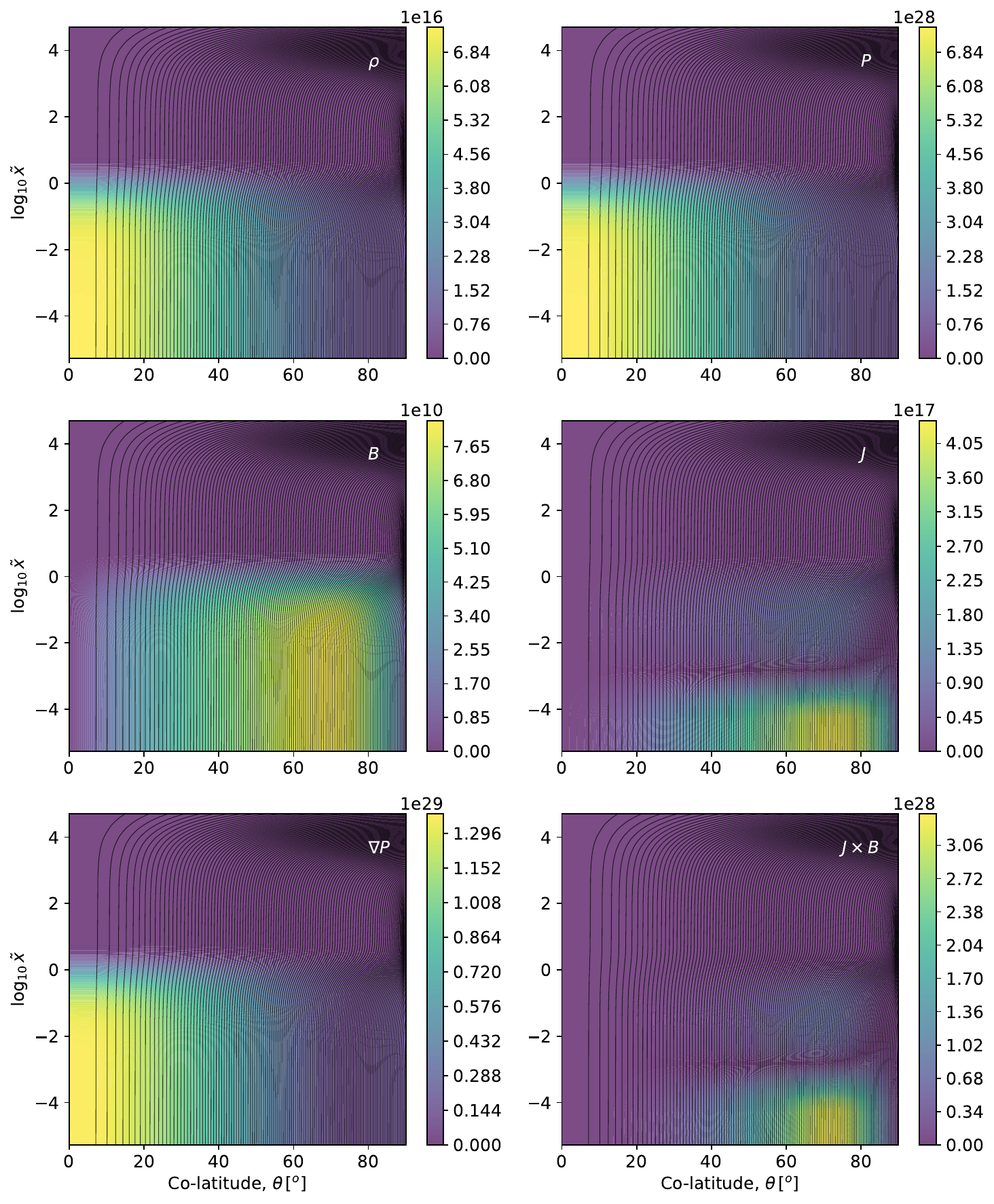}
\caption{Hydromagnetic configuration for a mountain with cross-field mass transport and $b=10$, $M_{\rm a} = 10^{-5}M_\odot$. Panels display the density $\rho$ (top left); pressure $p$ (top right); magnetic field strength $B$ (bottom left); and current density $|\mathbf{j}| = |\nabla\times\mathbf{B}|/\mu_0$ (bottom right), with $B_\ast = 10^8~\mathrm{T}$, $R_\ast = 10^4~\mathrm{m}$. Contours of $\psi$ are overlaid.}
\label{fig:combined_hydromag_structure}
\end{figure*}

A richer view of the hydromagnetic structure of the mountain is presented in Figure \ref{fig:combined_hydromag_structure}. Contours of the density $\rho$ and pressure $p$ are plotted in the top left and right panels (coloured contours) respectively, each overlaid with contours of $\psi$ (black contours). The magnetic field strength $|\mathbf{B}|$ and current density $|\mathbf{j}|$ are plotted in the middle panels, which confirm that screening currents, confined within a scale height $\sim x_0$ of the stellar surface, escape the polar cap by cross-field mass transport and spread to the equator. The pressure gradient $|\nabla p|$ and Lorentz force density $|\mathbf{j}\times\mathbf{B}|$ are plotted in the bottom panels. They balance one another to maintain equilibrium, but the magnitude of the force densities are lower when compared to those without cross-field mass transport.
The hydromagnetic structure is manifestly different to the structure for $b=10$ calculated by \citet{2004MNRAS.351..569P} (see Figure 4 in the latter reference), because cross-field mass transport supplants flux freezing. Indeed, $\psi$ in Figure \ref{fig:psi_ma_1e-5} for $b=10$ (bottom panel) is more akin to $\psi$ for $b=3$, which corresponds to a larger polar cap radius, consistent with equatorward spreading through cross-field mass transport. We plot the flux surfaces for $M_{\rm a}/M_\odot = 10^{-5}$ and $b=3$ in the top panel of Figure \ref{fig:psi_ma_1e-5}, again using the fit \eqref{eq:parameter_fit_sech} to generate $dM/d\psi$, highlighting the broad similarities after cross-field mass transport between the two panels.

\section{Astrophysical observables}{\label{sec:astro_observables}}
In this section we study the observational implications of our model, investigating the scalings of the magnetic dipole moment $|\mathbf{m}_\mathrm{d}|$ and the mass quadrupole moment (equivalently the mass ellipticity $\epsilon$) as functions of accreted mass $M_{\rm a}$. The dipole moment is measured directly through pulsar timing and indirectly through arguments involving magnetocentrifugal equilibrium and binary evolution \citep{1986ApJ...305..235T, 1991PhR...203....1B, 1995xrbi.nasa..495B,1995A&A...297L..41V}. The mass quadrupole moment is a source of continuous gravitational waves \citep{1980RvMP...52..299T, 2008MNRAS.389..839W, 2023MNRAS.521.2103L,2023LRR....26....3R}. Gravitational wave searches with audio-band, long-baseline, terrestrial interferometers have placed upper limits on $\epsilon$ for neutron stars with an accretion history \citep{2007PhRvD..76h2001A,2015PhRvD..91f2008A,2017PhRvD..95l2003A, 2017ApJ...847...47A,2019PhRvD.100l2002A,2020ApJ...902L..21A,2020PhRvD.102b3006M,2021ApJ...906L..14Z,2022ApJ...929L..19C,2022PhRvD.105b2002A, 2023APh...15302880W, 2025PhRvD.111h4040V,2025MNRAS.tmp.2069P}, and there are good prospects for obtaining improved $\epsilon$ upper limits in the near future \citep{2023LRR....26....3R}.

In this section, we calculate $|{\bf m}_{\rm d}|$ and $\epsilon$ to cover the astrophysically relevant regime $M_{\rm a} \lesssim 10^{-1} M_\odot$, consistent with the maximum $M_{\rm a}$ inferred from the ages and evolutionary histories of typical accreting systems \citep{1986ApJ...305..235T, 2008MNRAS.389..839W}. To do so, we exploit the general, two-parameter formulas \eqref{eq:parameter_fit_sech} and \eqref{eq:parameter_fit_f_func} for the mass-flux distribution, which are calibrated in Section \ref{sec:hydromag_structure_of_mtn} and Appendix \ref{sec:app_C_dmdpsi_fit} in the computationally feasible regime $M_{\rm a} \leq 10^{-5} M_\odot$. We emphasise that mountain growth in the extrapolated regime $10^{-5} \lesssim M_{\rm a}/M_\odot \lesssim 10^{-1} M_\odot$ is accompanied by the formation of magnetic bubbles \citep{1989ApJ...345.1034K, 2004MNRAS.351..569P}, which are not described by \eqref{eq:parameter_fit_sech} and \eqref{eq:parameter_fit_f_func}, so the results should be regarded as indicative rather than conclusive. An in-depth study of magnetic bubbles forming over the accretion time-scale $\sim M_{\rm a} / \dot{M}_{\rm a}$ requires expensive, time-dependent, MHD simulations with a prohibitively wide dynamic range and lies outside the scope of this paper.

\subsection{Magnetic dipole moment}\label{sec:astro_obs/dipole_moment}
\begin{figure}
\centering
\begin{subfigure}{\columnwidth}
\includegraphics[width=\columnwidth]{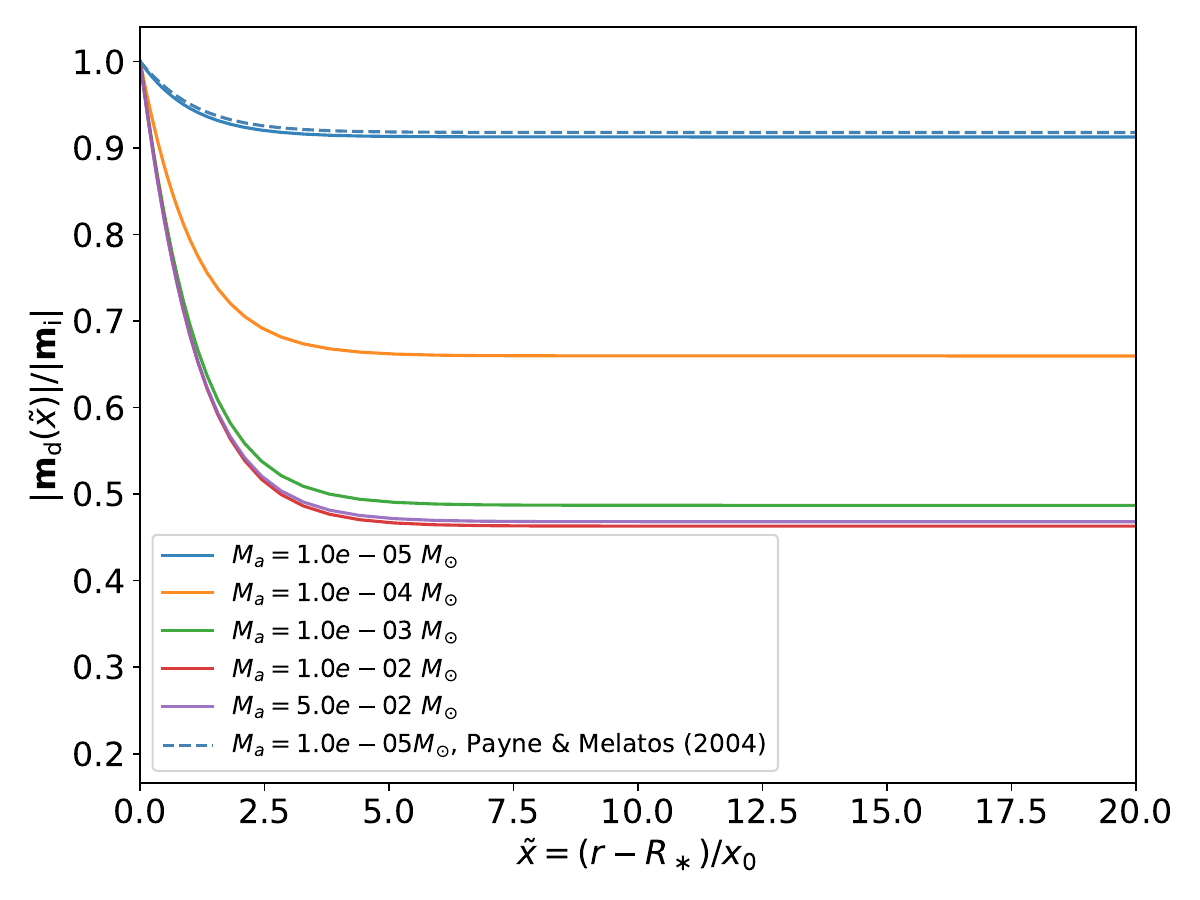}
\end{subfigure}

\begin{subfigure}{\columnwidth}
\includegraphics[width=\columnwidth]{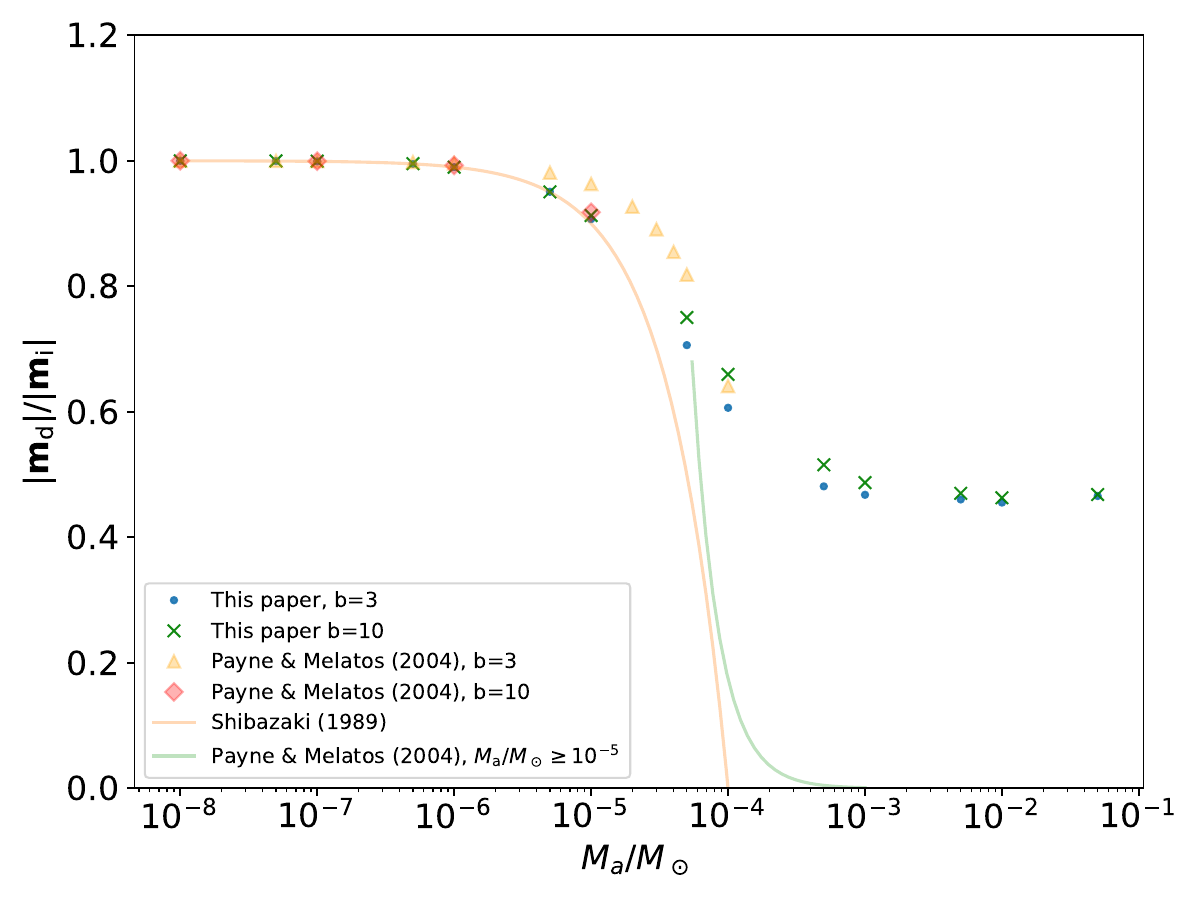}
\end{subfigure}
\caption{Reduction of the magnetic dipole moment due to accretion. Top panel plots the normalised dipole moment $|\mathbf{m}_{\rm d}|/|\mathbf{m}_{\rm i}|$ as a function of altitude above the stellar surface, $\tilde{x} = (r-R_\ast)/x_0$.  Solid curves are the reduced dipole moment for the functional fit equation \eqref{eq:parameter_fit_sech}, for $M_{\rm a}/M_{\odot} = 10^{-5}$, $10^{-4}$, $10^{-3}$, $10^{-2}$, and $5\times10^{-2}$, for $b=10$. Dashed curve corresponds to $M_{\rm a}/M_\odot=10^{-5}$ from \citet{2004MNRAS.351..569P}, also for $b=10$. The minimum value of the dipole moment occurs at $M_{\rm a}=5\times10^{-2}M_\odot$ and is $|\mathbf{m}_{\rm d}|/|\mathbf{m}_{\rm i}| = 0.46$. The bottom panel plots the normalised dipole moment $|\mathbf{m}_{\rm d}|/|\mathbf{m}_{\rm i}|$ as a function of accreted mass $M_{\rm a}/M_\odot$ for $b=3$ (blue points) and $b=10$ (green crosses), plotted alongside the scaling $|\mathbf{m}_{\rm d}|/|\mathbf{m}_{\rm i}| = (1 + M_{\rm a}/10^{-4}M_\odot)^{-1}$ of \citet{1989Natur.342..656S} (orange curve), as well as the empirical fit of \citet{2004MNRAS.351..569P}, $|\mathbf{m}_{\rm d}|/|\mathbf{m}_{\rm i}| = (M_{\rm a}/4.6\times10^{-5}M_\odot)^{-2.25\pm0.22}$ for $M_{\rm a} \gtrsim 5\times 10^{-5}M_\odot$ (green curve). The effect of mass transport is to prevent the dipole from being reduced. Instead it plateaus, as the magnetic field relaxes back towards a more dipolar configuration.}
\label{fig:dipole_moment}
\end{figure}

The magnetic dipole moment measured at radius $r$,
\begin{equation}
    |\mathbf{m}_{\rm d}| = \frac{3r^3}{4}\int_{-1}^{1} d(\cos\theta)\cos\theta B_r(r,\theta),
    \label{eq:dipole_moment_definition}
\end{equation}
is reduced by screening currents in the mountain, which deform the magnetic field. Hence, for any given hydromagnetic equilibrium with $\psi(R_\ast,\theta) \propto \sin\theta$ corresponding to an undisturbed dipole at the stellar surface, ${\bf m}_{\rm d}$ decreases from its pre-accretion value ${\bf m}_{\rm i}$ at $r=R_\ast$ to an asymptotic, reduced (i.e.\ screened) value at $r \gg R_\ast$ (e.g.\ at the edge of the simulation volume). To illustrate this, the top panel of Figure~\ref{fig:dipole_moment} displays the normalised dipole moment $|\mathbf{m}_{\rm d}|/|\mathbf{m}_{\rm i}|$ as a function of distance from the stellar surface, $\Tilde{x}=(r-R_\ast)/x_0$, for $M_{\rm a} \leq 5\times 10^{-2} M_\odot$ with cross-field mass transport implemented (solid curves).\footnote{Numerical errors destabilise the solver for $M_{\rm a} / M_\odot \gtrsim 5\times 10^{-2}$, primarily stemming from difficulties in resolving closely bunched flux surfaces near the equatorial plane, where the field line curvature is greatest.} The dipole moment is calculated once all unstable regions have been nullified by mass transport. We see that $|\mathbf{m}_{\rm d}|$ decreases monotonically with $M_{\rm a}$, and attains an asymptotic value at $r\gg R_\ast$ for every plotted $M_{\rm a}$ value. The top panel of Figure \ref{fig:dipole_moment} plots $|\mathbf{m}_{\rm d}|$ for $b=10$ alone, but asymptotic values for $b=3$ are within $\sim3$ per cent and so are omitted for clarity. Cross-field mass transport has a minor effect on the dipole moment reduction for $M_{\rm a} \leq 10^{-5} M_\odot$, with $|\mathbf{m}_{\rm d}|/|\mathbf{m}_{\rm i}|$ differing by $\leq0.6$ per cent at $r \gg R_\ast$ with and without mass transport, e.g.\ compare the solid and dashed curves for $M_{\rm a} = 10^{-5} M_\odot$ in the top panel of Figure \ref{fig:dipole_moment}. 

Cross-field mass transport hinders the screening of the magnetic dipole moment for $M_{\rm a} \geq 10^{-5} M_\odot$. This is natural: magnetic flux surfaces do not migrate equatorward as far as they do under flux-freezing conditions, more magnetic flux emerges near the pole, and $|\mathbf{m}_{\rm d}|$ is relatively greater. As the accreted mass approaches the astrophysically relevant regime $M_{\rm a} \sim 10^{-2} M_\odot$ , the dipole moment plateaus to $|\mathbf{m}_{\rm d}|/|\mathbf{m}_{\rm i}| \approx 0.46$, as shown in the top panel of Figure \ref{fig:dipole_moment}. An incremental unit of mass accreted at this stage represents a smaller fraction of the mountain's total mass compared to earlier accretion, and the change in the mountain profile is relatively smaller than when an equal unit of mass is accreted earlier in the quasistatic assembly. Hence, changes in the screening currents are similarly relatively smaller, and $\mathbf{m}_{\rm d}$ plateaus, instead of monotonically decreasing.

Previous authors fitted $|\mathbf{m}_{\rm d}|/|\mathbf{m}_{\rm i}|$ -- without cross-field mass transport -- with a power-law scaling. \citet{2004MNRAS.351..569P} reported a fit of the form $|{\bf m}_{\rm d}| / |{\bf m}_{\rm i}| \propto (M_{\rm a}/4.6\times10^{-5}M_\odot)^{-2.25\pm0.22}$, whilst \citet{1989Natur.342..656S} introduced the classic, astrophysically motivated scaling $|{\bf m}_{\rm d}| / |{\bf m}_{\rm i}| \propto (1+M_{\rm a}/M_{\rm c})^{-1}$, applied subsequently in many contexts. This stands in contrast to the findings here, where cross-field mass transport results in the magnetic dipole moment saturating to a constant value. We plot $|\mathbf{m}_{\rm d}|/|\mathbf{m}_{\rm i}|$ asymptotically as a function of $M_{\rm a}$ in the bottom panel of Figure \ref{fig:dipole_moment}, for $b=3$ (blue points) and $10$ (orange crosses). For comparison, we also plot the corresponding numerical results from \citet{2004MNRAS.351..569P} for $b=3$ (green triangles) and $10$ (red diamonds), as well as the aforementioned empirical fits of \citet{1989Natur.342..656S} (green) and \citet{2004MNRAS.351..569P} (red), the latter for $M_{\rm a}/M_\odot \gtrsim 10^{-5}$.

Previous analyses suggest that screening by a magnetic mountain may explain the relatively low $|{\bf m}_{\rm d}|$ of millisecond pulsars \citep{1974SvA....18..217B,1983A&A...128..369H,1998ApJ...496..915B,2001PASA...18..421M,2004MNRAS.351..569P,2008MNRAS.386.1294V,2011MNRAS.417.2696P} and the ``bottom field'' of low-mass X-ray binaries \citep{1998A&A...330..195Z, 2002MNRAS.332..933C, 2006MNRAS.366..137Z}. None of these previous, ideal-MHD analyses incorporate cross-field mass transport as implemented through the semi-analytic recipe proposed by \citet{2020JPlPh..86f9002K}. It is therefore interesting to ask whether some other process beyond diamagnetic screening, such as Ohmic dissipation, is required to explain the available $|{\bf m}_{\rm d}|$ data from a population standpoint. At the time of writing, it is hard to say for sure. Certainly, cross-field mass transport acts generically to hinder diamagnetic screening, as confirmed by the simulations in this paper, but the degree to which this happens remains challenging to quantify. For one thing, the scalings in Figure \ref{fig:dipole_moment} neglect the formation of magnetic bubbles, as noted above \citep{1989ApJ...345.1034K,2004MNRAS.351..569P}. It may be argued that bubble formation hinders screening even more, of course, but such a claim needs to be tested rigorously. Moreover, the \citet{2020JPlPh..86f9002K} recipe for cross-field mass transport is not the only possible recipe, although it is well motivated physically with reference to analogous systems in laboratory plasmas. Other recipes may give different answers; a firm resolution of these issues must await refined, time-dependent, nonideal-MHD simulations in the future, which face steep challenges due to the wide dynamic range of the problem (i.e.\ from the fast Alfv\'{e}n time-scale to the slow accretion time-scale), as noted above. 
\subsection{Mass quadrupole moment}\label{sec:astro_obs/quadrupole}
\begin{figure}
    \centering
    \includegraphics[width=\columnwidth]{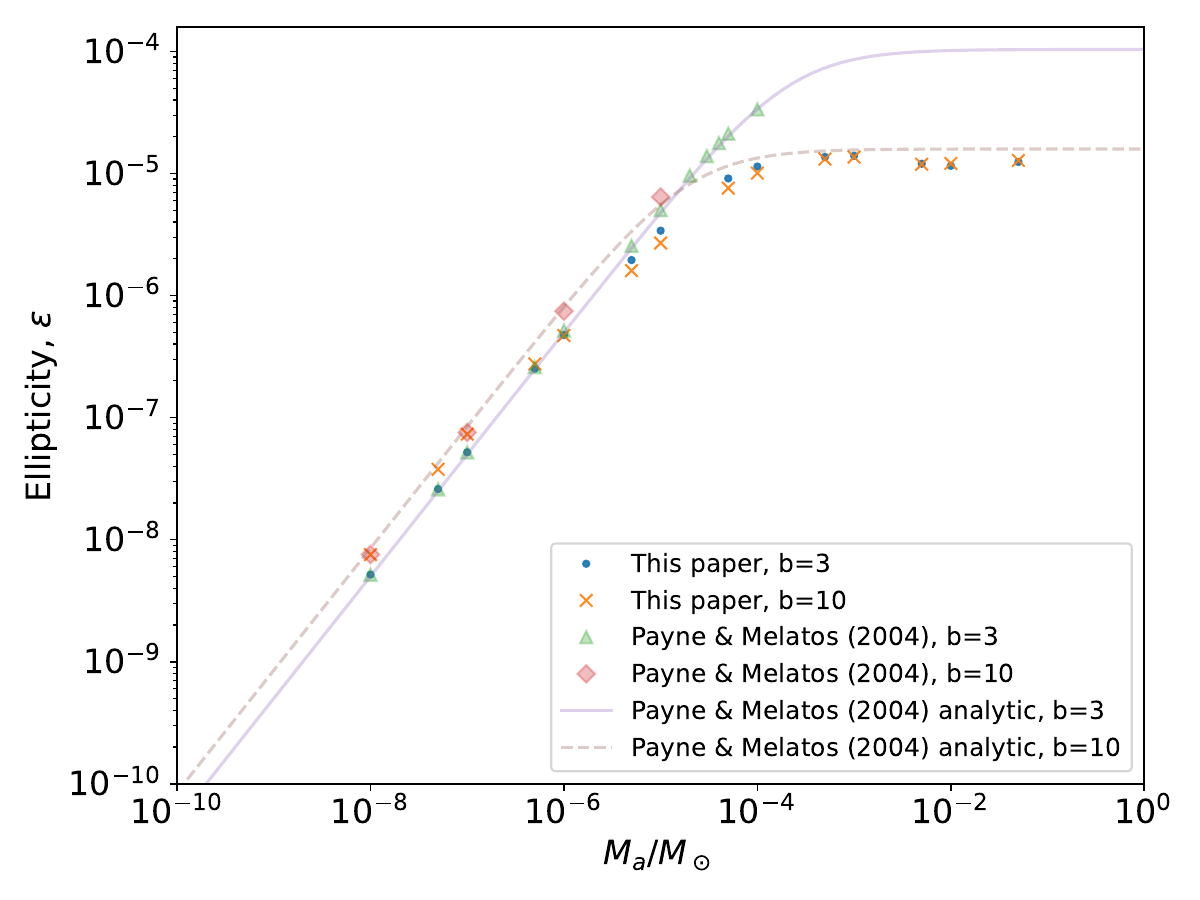}
    \caption{Mass ellipticity $\epsilon$ as a function of accreted mass $M_{\rm a}$. Analytic curves for $b=3$ (solid) and $b=10$ (dashed) from \citet{2004MNRAS.351..569P}. Numerical results using $dM/d\psi$ from equation \eqref{eq:dmdpsi_pm04} for $b=3$ (triangles), $b=10$ (diamonds), and numerical results using equation \eqref{eq:parameter_fit_sech} for $b=3$ (points) and $b=10$ (crosses).}
    \label{fig:ellipticity_0}
\end{figure}

The mass ellipticity is defined as $\epsilon = |I_1 - I_3|/I_1$, where $I_1 = I_2 < I_3$ are the principal moments of inertia of a biaxial rotor, oriented such that $\mathbf{e}_1$ is directed along the magnetic dipole moment $\mathbf{m}_{\rm d}$. We calculate
\begin{equation}
    \epsilon = \pi I_0^{-1}\int_{-1}^{1} d(\cos\theta)\int_0^{R_\ast}dr ~r^4(3\cos^2\theta - 1)\rho(r,\theta)
\end{equation}
as a function of $M_{\rm a}$ and display the results as points in Figure \ref{fig:ellipticity_0} for $b=3$ (blue points) and 10 (orange crosses). The ellipticity increases monotonically with $M_{\rm a}$ as expected. For accreted masses $\leq 5\times10^{-7}M_\odot$, the ellipticity of the mountain with $b=10$ is approximately 40 per cent larger than the mountain with $b=3$. For $M_{\rm a} \geq 10^{-5} M_\odot$, $\epsilon$ for $b=10$ becomes smaller than for $b=3$. For $M_{\rm a}=10^{-4}M_\odot$, $\epsilon$ for $b=3$ and 10 saturates to the value $\epsilon\approx10^{-5}$.

There are several differences between these results and previous work, most notably \citet{2004MNRAS.351..569P}. First, whilst \citet{2004MNRAS.351..569P} observed $\epsilon$ increasing monotonically with $M_{\rm a}$, they found that $\epsilon$ for $b=3$ (green triangles in Figure \ref{fig:ellipticity_0}) becomes larger than for $b=10$ (red diamonds)  for $M_{\rm a} \gtrsim 1.4\times 10^{-5}M_\odot$ cf.\ $M_{\rm a} \gtrsim 5\times10^{-7}M_\odot$ in this work. Second, \citet{2004MNRAS.351..569P} observe a continued monotonic increase in $\epsilon$ for both $b=3$ and 10, before running into numerical constraints. Here, we are able to accrete sufficient mass in order to confirm that saturation does occur. Third, we observe saturation to $\epsilon\approx 10^{-5}$. Cross-field mass transport leads to mountains with differing initial profiles resembling one another once stabilised. 
\section{Conclusions}{\label{sec:6}}
In this paper, we implement the cross-field mass transport recipe proposed by \citet{2020JPlPh..86f9002K} to calculate the evolution of a polar magnetic mountain as it grows through accretion. The calculation is based on a self-consistent iterative scheme. It combines a semi-analytic mass transport prescription, which nullifies the ideal Schwarzschild instability locally, and a numerical solver for the Grad-Shafranov equation, which solves for the hydromagnetic equilibrium globally, given the mass-flux distribution $dM/d\psi$ resulting from mass transport. Table~\ref{tab:comparison} summarises the key differences in input physics and results between \citet{2004MNRAS.351..569P} and this work. In keeping with previous work, we find that hydromagnetic instabilities occur locally but do not demolish the mountain globally. Instead, the mountain adjusts by shuffling a modest fraction of its mass equatorward across magnetic flux surfaces, so that it remains marginally stable on every flux surface. This conclusion holds whether the mountain is assembled in a single shot or through a quasistatic sequence of accretion followed by adjustment, although the exact details of the final state depend somewhat on the path (Section \ref{sec:path_dependence}). Quasistatic assembly is more realistic astrophysically, because the Alfv\'{e}nic instability growth time-scale is much shorter than the accretion time-scale (Section \ref{subsec:quasistatic_approx}).

\begin{table*}
    \centering

    \caption{Comparison of input physics and results between \citet{2004MNRAS.351..569P} and this work.}
    \label{tab:comparison}
    \begin{tabular}{lll}
        \hline
        & \citet{2004MNRAS.351..569P} & This work \\
        \hline
        \multicolumn{3}{l}{\textit{Input physics}} \\
        \hline
        Equation of state & Isothermal, $p = c_s^2\rho$ & Isothermal, $p = c_s^2\rho$ \\
        Flux freezing & Enforced (ideal MHD) & Broken locally by Schwarzschild instability \\
        Mass-flux distribution $dM/d\psi$ & Exponential always, see~\eqref{eq:dmdpsi_pm04} & Exponential initially; modified by cross-field transport \\
        Instability & None & Schwarzschild instability; nullified iteratively \\
        Surface boundary condition & Line-tying, $\psi(R_*, \theta) = \psi_d(R_*, \theta)$ & Line-tying, $\psi(R_*, \theta) = \psi_d(R_*, \theta)$ \\
        Outer boundary condition & Radial, $\partial\psi/\partial r=0$ & Constant dipole moment \citep{2023MNRAS.526.2058R} \\
        Maximum $M_a$ & $\sim 3\times10^{-5}M_\odot$ (solver halts) & $\sim 10^{-2}M_\odot$ via empirical fit ~\eqref{eq:parameter_fit_sech}-\eqref{eq:parameter_fit_f_func} \\
        \hline
        \multicolumn{3}{l}{\textit{Results}} \\
        \hline
        $F(\psi)$ & Self-consistent via~\eqref{eq:Fpsi_01} & Self-consistent via~\eqref{eq:Fpsi_01}; modified by cross-field transport \\
        Screening currents & Confined to polar cap & Spread to equator \\
        Dipole moment $|\mathbf{m}_d|/|\mathbf{m}_i|$ & Power-law decay, $\propto (M_a/4.6\times10^{-5}M_\odot)^{-2.25\pm0.22}$ & Plateaus to $\approx 0.46$ for $M_a \gtrsim 10^{-2}M_\odot$ \\
        Mass ellipticity $\epsilon$ & Monotonically increasing; no saturation observed & Saturates to $\epsilon \approx 10^{-5}$ for $M_a \gtrsim 10^{-4}M_\odot$ \\
        
        \hline
    \end{tabular}
\end{table*}

We find that cross-field mass transport modifies the hydromagnetic structure of the mountain, allowing screening currents to escape the polar region to spread equatorwards. Pressure gradients and Lorentz force densities are smaller when compared to a mountain without cross-field mass transport. We show that the hydromagnetic structure of the mountain depends on whether the mountain is assembled quasistatically in incremental steps, or in ``one shot''. The mass-flux distribution in the former case is relatively steeper at the pole when compared to the latter case ($dM/d\psi$ at $\psi/\psi_\ast=0$ is 38 per cent larger than in the ``one shot'' case), though they are similar beyond the polar cap.

We find that the mass-flux distribution dictating the structure of the mountain may be described by a simple two parameter function, parameterised by accreted mass $M_{\rm a}$ and $b=\psi_\ast/\psi_{\rm a}$. We use the function to study the hydromagnetic structure up to the astrophysically relevant regime $10^{-5} \leq M_{\rm a}/M_\odot\leq 10^{-2}$, which cannot be accessed directly by simulations due to computational cost. We find that the inclusion of mass transport results in the hydromagnetic structure of the mountain becoming effectively independent of $b$, and that the magnetic flux surfaces partly relax back towards the undistorted, pre-accretion dipole. We calculate the reduced magnetic dipole moment $|\mathbf{m}_{\rm d}|$ and find that it is screened by mass accretion, but cross-field mass transport reduces the effectiveness of the screening for accreted masses $M_{\rm a} /M_\odot\geq 10^{-5}$. The asymptotic normalised dipole moment for $r\gg R_\ast$ plateaus to $|\mathbf{m}_{\rm d}|/|\mathbf{m}_{\rm i}|=0.46$ for $M_{\rm a}/M_\odot\leq 10^{-2}$. Similarly, the mass ellipticity saturates to $\epsilon \approx 10^{-5}$ for $M_{\rm a}/M_\odot \leq 10^{-2}$. The saturation of $\epsilon$ observed here has important consequences for the generation of continuous gravitational waves \citep{2023LRR....26....3R}, and will be studied in a future paper.

The regime in this paper is relevant to two observationally motivated scenarios: old neutron stars in low mass X-ray binaries, which accumulate $M_{\rm a} \gtrsim 10^{-2} M_\odot$ over $10^6 \, {\rm yr}$ at $\dot{M}_{\rm a}\sim10^{-8} M_\odot \rm\, yr^{-1}$; and young neutron stars experiencing accelerated accretion rates from a supernova fallback disc, which may deposit a comparable mass in $\lesssim 10^3 \, {\rm s}$. The latter scenario was analysed in detail by \citet{2014ApJ...794..170M} and is reviewed briefly in Appendix \ref{sec:app_E_newborn}. Although $\dot{M}_{\rm a}$ during fallback is $\gtrsim 10^{10}$ times higher than in a typical low-mass X-ray binary, the mountain building mechanism is the same, as (16) depends on $M_{\rm a}$ explicitly but not $\dot{M}_{\rm a}$. Material is funnelled to the polar cap of a protomagnetar by the strong magnetic field, dragging the field equatorwards, burying the magnetic dipole and increasing the mass quadrupole. The mass quadrupole is substantially larger than those found in this paper, scaling as $\epsilon = 3\times10^{-3}(|\mathbf{m}_{\rm i}|/10^{23}\,\mathrm{T\,m^3})^2$, owing to the stronger magnetic field. The burial of the magnetic dipole by the mountain suppresses the onset of the propeller phase regulating mass accretion onto the protomagnetar, which in turn assists black hole formation \citep{2011ApJ...736..108P,2012ApJ...761...63P,2014ApJ...794..170M}.

It is interesting to ask how the X-ray luminosity of an accreting neutron star affects a mountain built on the surface. X-ray luminosity is a proxy for the mass accretion rate $\dot{M}_{\rm a}$. It affects the time to build a mountain. In contrast, it does not affect the mountain structure directly, e.g.\ the solution of \eqref{eq:GS_eqn_03} depends on $M_{\rm a}$ explicitly but not $\dot{M}_{\rm a}$. However, $\dot{M}_{\rm a}$ does affect the mountain structure indirectly in two ways. First, it helps to set the magnetospheric radius $R_{\rm a}$, which sets in turn the geometry of the accretion column and the size of the polar cap (quantified by $b=\psi_\ast/\psi_{\rm a} = R_{\rm a}/R_\ast$ in this paper). 
Different accretion rates affect how the magnetic field threads the accretion disc, and hence determine the form of the mass-flux distribution. Second, $\dot{M}_{\rm a}$ and hence the X-ray luminosity affect the rate of surface heating and its variation with latitude and longitude, which in turn affects the mountain structure through the equation of state \citep{2011MNRAS.417.2696P} and thermal conduction \citep{2019MNRAS.484.1079S}. Specifically, $\epsilon$ decreases from $\sim 10^{-4}$ to $\sim10^{-6}$ when one transitions from an isothermal to a polytropic equation of state (non-relativistic degenerate neutrons), while thermal conduction promotes the flow of matter polewards, increasing ellipticity compared to models with no conduction.

We make several simplifying assumptions in this paper to keep the problem tractable computationally, which should be tested in future work and modified where necessary. (i) We consider axisymmetric equilibria. It is known that the Schwarzschild instability behaves differently in three dimensions than in two, as does the undular submode of the Parker instability, although there are indications that the final outcome is similar, in the sense that the mountain is not disrupted globally in either setting \citep{1992PASJ...44..167M, 2008MNRAS.386.1294V}. (ii) We adopt an isothermal equation of state. A polytropic equation of state is a natural extension to this work, as in \citet{2011MNRAS.417.2696P}. (iii) We assume that the magnetic field lines are tied rigidly to the surface of the crust, and that no sinking occurs. Sinking has been considered by other authors previously \citep{2002MNRAS.332..933C, 2010MNRAS.402.1099W}. (iv) We do not consider the impact of type-II superconductivity in the interior \citep{2011MNRAS.410..805G, 2012MNRAS.419..732L, 2021PASA...38...43S}, even though it is known that superconductivity expels the toroidal field from the core to the crust, modifying the hydromagnetic geometry at the mountain base. (v) We assume that the accreted mass-flux distribution follows the exponential profile \eqref{eq:dmdpsi_pm04}, rather than ring-shaped profiles and their variants, which arguably capture more accurately the interaction between the accretion column on open magnetic field lines and the inner edge of the accretion disk \citep{2013MNRAS.430.1976M,2013MNRAS.435..718M,2025MNRAS.541.3280Y}. (vi) We neglect the formation of magnetic bubbles \citep{1989ApJ...345.1034K,2004MNRAS.351..569P} for $M_{\rm a} \gtrsim 10^{-5} M_\odot$, an assumption whose consequences are discussed in detail in Section \ref{sec:astro_observables}. (vii) We ignore higher order multipoles, even though phase-resolved X-ray spectroscopic observations of the surface temperature distribution have revealed that the magnetic field cannot be that of a centred dipole \citep{2019ApJ...878L..22R,2019ApJ...887L..21R}. Grad Shafranov equilibria for fields with higher order multipoles have been studied by \citet{2020MNRAS.499.3243S} and \citet{2025MNRAS.541.3280Y} and are a natural extension to the initial investigation undertaken here.
(viii) We neglect the effect of stellar rotation, which introduces centrifugal and Coriolis force densities to the momentum equation \eqref{eq:momentum_cons_01}. We show in Appendix \ref{sec:app_D_coriolis} that the inclusion of such terms results in small corrections to the results reported here, which may be neglected. Ultimately, a full dynamic evolution of the mountain using a three-dimensional, non-ideal MHD solver with sufficient resolution and dynamic range will be needed to investigate fully the stability of an accreted magnetic mountain, instead of the simplified study presented here.

\section*{Acknowledgements}
We wish to thank Arthur Suvorov, Maxim Priymak, Matthias Vigelius and Donald Payne for sharing the original Grad-Shafranov code which this paper extends, as well as assistance in setting it up. This work received financial support from the Australian Research Council Centre of Excellence for Gravitational Wave Discovery (OzGrav), project number CE230100016. Pedro H.B. Rossetto acknowledges support from the São Paulo State Funding Agency (FAPESP). This study was financed, in part, by the São Paulo Research Foundation (FAPESP), Brasil. Process Number 2024/21854-2. This work used computational resources of the OzSTAR national facility at Swinburne University of Technology. OzSTAR is funded by Swinburne University of Technology and also the National Collaborative Research Infrastructure Strategy (NCRIS). This work made use of the Python programming language and the NumPy and SciPy libraries.

\section*{Data Availability}
The data underlying this article will be shared on reasonable request to the corresponding author.


\bibliographystyle{mnras}
\bibliography{references} 




\appendix

\section{Cross-field mass transport schemes}\label{sec:app_A_mass_transport}
In this appendix, we justify physically the mass transport scheme \eqref{eq:delta_mass_transport_01} and \eqref{eq:update_dmdpsi_accretion_delta_m_00} used to stabilize flux surfaces by nullifying density gradients across two or more adjacent flux tubes. We distinguish between nullifying density gradients locally (at a point on a flux surface) and globally (everywhere along a flux surface) and show that local and global adjustments are equivalent under certain physical conditions.

\subsection{Equalising mass in adjacent straight flux tubes}\label{sec:app_A_equalising_mass_scheme}
To start we consider a region sufficiently close to the stellar surface, such that the surface may be taken as planar, and the curvature of the dipolar magnetic field lines is negligible. Consider a simple arrangement consisting of a single flux surface $\psi$ separating two flux tubes $A$ (with $\psi-\Delta\psi \leq \psi' \leq \psi$) and $B$ (with $\psi \leq \psi' \leq \psi+\Delta\psi$). The flux tubes are rectangular in cross-section and stand perpendicular to the $x$-$y$ plane in a local Cartesian coordinate system, such that the magnetic field ${\bf B}$ is aligned with the $z$-axis, and the flux surface $\psi$ occupies the $y$-$z$ plane. Each flux tube contains plasma in hydrostatic equilibrium, satisfying an isothermal equation of state (the argument holds for other equations of state too). We assume a linearised gravitational potential $\Phi(z) = gz$, near the stellar surface. Then the mass density in flux tube $A$ is
\begin{equation}
    \rho_A(z) = \rho_A(0)\exp(-z/z_0),
    \label{eq:column_density_isothermal_eos_01}
\end{equation}
where $\rho_A(0)$ is the density at the base, and $z_0$ is the hydrostatic scale height. Equation \eqref{eq:column_density_isothermal_eos_01} also applies to flux tube $B$, after replacing the subscript $A$ with $B$. Integrating to height $z$, the mass $M_A(z)$ per unit cross-sectional area $S_A$ is given by
\begin{equation}
    \frac{M_A(z)}{S_A} =\rho_A(0)z_0\left[1 - \exp(-z/z_0)\right]
    \label{eq:mass_equalisation_formula_1}
\end{equation}
for flux tube $A$, and an analogous formula applies to flux tube $B$. Now, to nullify the density gradient across the flux surface $\psi$, one must equalise the densities in each flux tube for all $z$. Equation \eqref{eq:column_density_isothermal_eos_01} then implies $\rho_A(0) = \rho_B(0)$, and in turn \eqref{eq:mass_equalisation_formula_1} implies
\begin{equation}
    \frac{M_A(z)}{S_A} = \frac{M_B(z)}{S_B}.
    \label{eq:mass_equalisation_formula_3}
\end{equation}

For $S_A = S_B$, equation (A5) implies $M_A(z)=M_B(z)$. That is, the density gradient is nullified at all $z$, when the masses in adjacent flux tubes are equal.
On the surface of an axisymmetric neutron star, a flux tube occupying the volume $\theta_A \leq \theta \leq \theta_A + \Delta\theta$ has $S_A = 2\pi R_\ast^2 [ \cos\theta_A - \cos(\theta_A + \Delta\theta) ]$. The numerical simulations in this paper adopt grid points spaced equally in $\cos\theta$, which implies $S_A=S_B$ and hence $M_A=M_B$ to nullify the density gradient (and hence hydromagnetic instabilities) between adjacent flux tubes.

On flux surfaces where the instability condition \eqref{eq:delta_criterion_00} is satisfied, \citet{2020JPlPh..86f9002K} proposed that the density gradient is nullified by the nonlinear evolution of the instability on the fast Alfv\'{e}n time-scale. To achieve this numerically, we equalise the mass in two adjacent flux tubes by moving an amount of mass $\Delta M$ between them. The total mass in the two flux tubes is $M_A + M_B$, so, after equalisation, the mass in each flux tube equals $(M_A+M_B)/2$, and $\Delta M = (M_B - M_A)/2$ moves out of flux tube $A$ into flux tube $B$.
For a given mass-flux distribution $(dM/d\psi)^{\rm (init)}$ before cross-field mass transport, the mass in flux tube $A$ equals
\begin{equation}
    M_A = \int_{\psi-\Delta\psi}^{\psi}d\psi'\left(\frac{dM}{d\psi'}\right)^{\mathrm{(init)}},
\end{equation}
and similarly for flux tube $B$. Upon substituting into the expression for $\Delta M$, one obtains
\begin{equation}
    \Delta M(\psi) = \frac{1}{2}\left[\int_{\psi}^{\psi + \Delta\psi} d\psi'\left(\frac{dM}{d\psi'}\right)^{\mathrm{(init)}} - \int_{\psi - \Delta\psi}^{\psi}d\psi' \left(\frac{dM}{d\psi'}\right)^{\mathrm{(init)}} \right],
    \label{eq:delta_mass_transport_02_appendixA}
\end{equation}
which matches equation \eqref{eq:delta_mass_transport_01} in Section \ref{subsec: cfmf}. 

When multiple adjacent flux surfaces are unstable, forming an unstable region, we equalise the mass across all flux tubes in the region in a similar manner. If there are $N$ unstable flux surfaces in the region $\psi_1\leq\psi\leq\psi_2$, the total mass in the corresponding $N+1$ flux tubes is
\begin{equation}
    M_{\rm total} = \int_{\psi_1-\Delta\psi}^{\psi_2}d\psi'\left(\frac{dM}{d\psi'}\right),
\end{equation}
where the lower terminal indicates that we move mass from the flux tube $\psi_1 - \Delta\psi$, across flux surface $\psi_1$, so the mass in $\psi_1 - \Delta\psi$ must be included in the sum. The mass moved from flux tube $\psi_i$ in the unstable region then becomes
\begin{equation}
    \Delta M(\psi) = \frac{M_{\rm total}}{N} - \int_{\psi_i - \Delta\psi}^{\psi_i}d\psi'\left(\frac{dM}{d\psi'}\right).
\end{equation}

\subsection{Alternative scheme: equalising mass locally}\label{sec:app_A2_local_mass_alt_scheme}
Instead of equalising the total mass in each flux tube, an alternative approach is to equalise mass locally, by considering two infinitesimal volumes adjacent to the section of the flux surface that is unstable according to equation \eqref{eq:delta_criterion_00}, and adjusting the mass in each of them such that the density on either side of the flux surface is equal. As in Appendix \ref{sec:app_A_equalising_mass_scheme}, the transported mass does not stay in the two  infinitesimal volumes; it quickly redistributes up and down along the flux tube to achieve hydrostatic equilibrium. The two schemes are equivalent physically, as explained below. In this paper, we adopt the scheme in Appendix \ref{sec:app_A_equalising_mass_scheme}. The scheme in Appendix \ref{sec:app_A2_local_mass_alt_scheme} is discussed to illustrate the kinds of alternatives which exist. 
\begin{figure}
    \centering
    \includegraphics[width=\columnwidth]{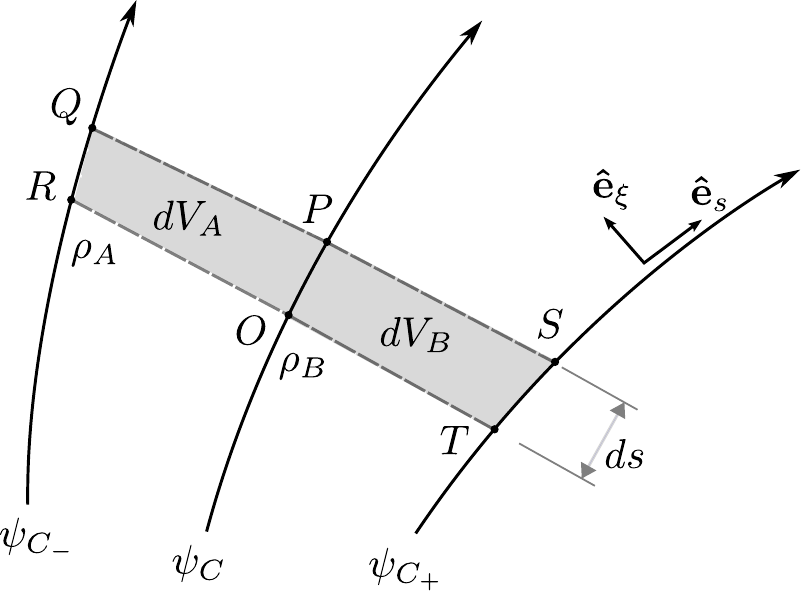}
    \caption{Geometry of cross-field mass transport: differential volumes \(dV_A\) and \(dV_B\) (shaded regions) between curved flux surfaces (solid curves), containing masses \(dm_A\) and \(dm_B\) respectively, for the calculation of the density gradient. If the segment \(OP\) is unstable, the mass \(\Delta m\) given by \eqref{eq:delta_M_03} is transported across \(OP\), such that the densities equalise. The labels are defined in Appendix \ref{sec:app_A_mass_transport}.}
    \label{fig:density_grad_01}
\end{figure}

Consider the configuration illustrated in \figureautorefname~\ref{fig:density_grad_01}. Two infinitesimal volume elements \(dV_A\) and \(dV_B\), containing masses \(dm_A\) and \(dm_B\) respectively, lie on either side of the flux surface \(\psi_C\). The local mass density in $dV_A$ is defined as
\begin{equation}
    \rho_A = \frac{dm_A}{dV_A},
    \label{eq:density_a_01}
\end{equation}
and an analogous formula defines $\rho_B$. We can then write
\begin{equation}
    dm_A = 2\pi\int_{\psi-d\psi}^{\psi} d\psi\int_{OP} ds\,r\sin\theta|\nabla\psi|^{-1}\rho_A.
    \label{eq:dm_01}
\end{equation}
For example, with reference to \figureautorefname~\ref{fig:density_grad_01}, we calculate \(dm_A\), where the flux surface \(\psi_C\) is unstable over the portion \(OP \subseteq C\), as follows: we first integrate along \(OP\) and the corresponding length \(QR\) along the flux surface \(\psi_{C-}\), then we integrate along \(\psi\) between \(\psi_{C-}\) and \(\psi_C\). An analogous formula applies for $dm_B$.
Similarly, the volume element \(dV_A\) is
\begin{equation}
    dV_A = 2\pi\int_{\psi-d\psi}^{\psi} d\psi\int_{OP} ds\, r\sin\theta|\nabla\psi|^{-1},
    \label{eq:dV_01}
\end{equation}
and analogously for $dV_B$. To equalise $\rho_A$ and $\rho_B$, we move an amount of mass $\Delta m$ from $A$ to $B$ satisfying
\begin{equation}
    \frac{dm_A - \Delta m}{dV_A} = \frac{dm_B + \Delta m}{dV_B}
    \label{eq:delta_M_01}
\end{equation}
and hence
\begin{align}
    \Delta m &= \frac{dm_A dV_B - dm_B dV_A}{dV_A + dV_B}\\
    &= \frac{\rho_A - \rho_B}{\frac{1}{dV_A} + \frac{1}{dV_B}}.
    \label{eq:delta_M_03}
\end{align}
If multiple segments of the flux surface are unstable, the total mass transported across \(\psi_C\) towards \(\psi_{C+}\) is found by summing the individual masses \(\Delta m\) for each of the unstable segments.

Following \citet{2020JPlPh..86f9002K}, we assume that the small-scale, turbulent, resistive dynamics of the Schwarzschild instability act to nullify the density gradient locally, as the instability develops non-linearly, until no further cross-field mass transport occurs. Therefore, the mass transport scheme described in Appendix \ref{sec:app_A_equalising_mass_scheme} corresponds to the final state of the transport scheme described in Appendix \ref{sec:app_A2_local_mass_alt_scheme}. It skips the intermediate steps of density gradients equalising locally at multiple unstable sites along a flux surface, to focus instead on the final, global outcome of the mass transport process.
\section{Numerical implementation}\label{sec:app_B_numerical_implementation}
\begin{figure}
    \centering
    \includegraphics[trim={5.5cm 6cm 4.5cm 4cm}, width=0.9\columnwidth]{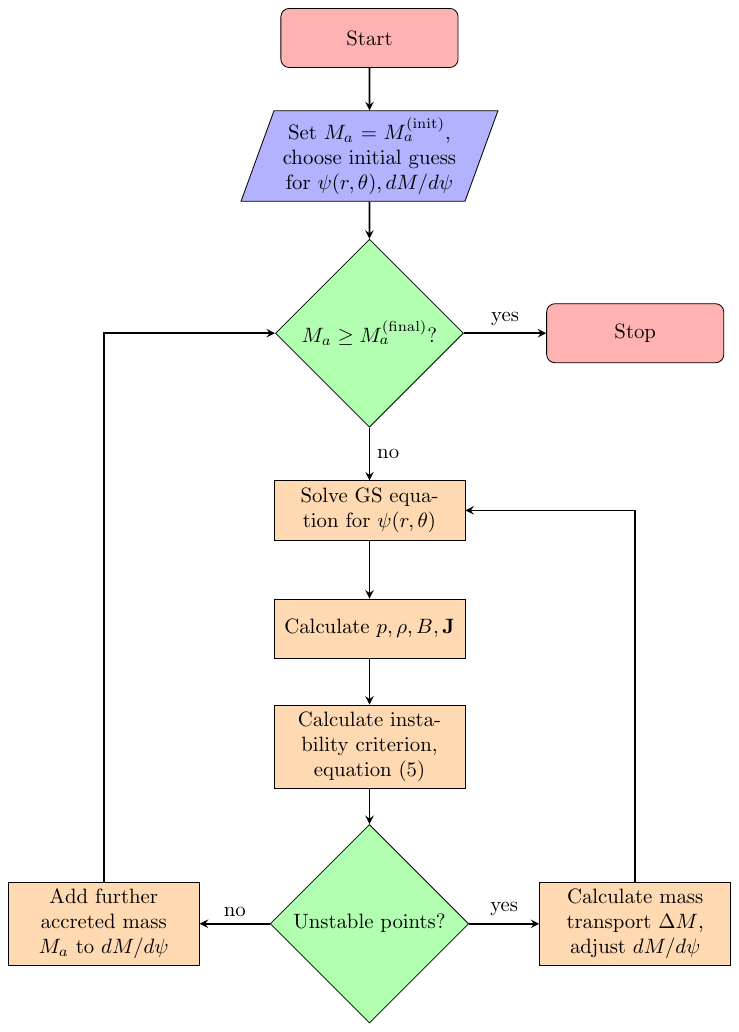}
    \caption{Flowchart of the numerical scheme. Yellow boxes indicate steps where a calculation is performed by the code, green diamonds indicate check conditions dictating the continuation or termination of an iterative loop, and red boxes indicate the start and end points.}
    \label{fig:cfd-flowchart}
\end{figure}

In this section we discuss the iterative numerical scheme used to solve the Grad-Shafranov equation and implement cross-field mass transport. A flowchart illustrating the steps involved in the numerical scheme is presented in \figureautorefname~\ref{fig:cfd-flowchart}.

\subsection{Grid and dimensionless variables}\label{sec:app_b1_grid_dim_vars}
The Grad-Shafranov equation \eqref{eq:GS_eqn_01} is solved in the region \(R_\ast\leq r \leq R_{\rm max}\) and \(0\leq \theta \leq \frac{\pi}{2}\), on a grid of dimension \((N_r, N_\theta)\). We transform to dimensionless coordinates \(\tilde{x} = (r - R_*)/x_0\) and \(\tilde{\mu} = \cos\theta\), and introduce the dimensionless variables \(\tilde{\psi} = \psi/\psi_0\), \(\tilde{M} = M/M_{\rm a}\), \(\tilde{F} = F/F_0\), \(\tilde{s} = s/x_0\), \(x_0 = c_s^2 R_*^2/GM_*\), and \(a = R_*/x_0\). Upon substitution, equations \eqref{eq:GS_eqn_03} and \eqref{eq:Fpsi_01} become
\begin{equation}
    \tilde{\Delta}_\ast\tilde{\psi} = -Q_0\tilde{F}^{\prime}(\tilde{\psi})e^{-\tilde{x}}
    \label{eq:GS_eqn_04}
\end{equation}
and
\begin{equation}
    \tilde{F}(\tilde{\psi}) = \frac{d\tilde{M}}{d\tilde{\psi}}\left[\int_{\mathcal{C}}d\tilde{s}(\Tilde{x}+a)(1-\Tilde{\mu}^2)^{1/2}|\Tilde{\nabla}\Tilde{\psi}|^{-1}e^{-\Tilde{x}}\right]^{-1}
    \label{eq:Fpsi_dimensionless_01}
\end{equation}
with \(Q_0 = \tilde{\mu}_0 x_0^4\), and the dimensionless Grad-Shafranov operator is defined as
\begin{equation}
    \tilde{\Delta}_\ast = \frac{1}{(\tilde{x} + a)^2(1-\tilde{\mu}^2)}\left[\frac{\partial^2}{\partial \tilde{x}^2} + \frac{1-\tilde{\mu}^2}{(\tilde{x} + a)^2}\frac{\partial^2}{\partial\tilde{\mu}^2}\right].
    \label{eq:GS_operator_dimensionless}
\end{equation}
We concentrate grid resolution near the stellar surface by scaling $\tilde{x}$ logarithmically as 
\begin{equation}
    \tilde{x}_1 = \log(\tilde{x} + e^{-L_x}) + L_x
    \label{eq:log_x_coordinates}
\end{equation}
and pick $L_x$ such that there are several points per scale height $x_0$. In order to perform the integration along field lines, we use the Python package \texttt{contourpy} to extract the interpolated coordinates \((\tilde{\mu}, \tilde{x})\) of each field line. We parameterise the length \(s\) along the field line, viz.
\begin{equation}
    \tilde{s} = \int_{\mathcal{C}}d\tilde{s},
\end{equation}
with
\begin{equation}F
    d\tilde{s} = \sqrt{d\tilde{x}^2 + \frac{(\tilde{x} + a)^2}{1 - \tilde{\mu^2}} d\tilde{\mu}^2}.
\end{equation}
Gradient terms, such as \(|\tilde{\nabla}\tilde{\psi}|\) in equation~\eqref{eq:Fpsi_dimensionless_01}, are calculated using second-order central differences at interior points, and second-order forwards and backwards differences at the boundaries.
\subsection{Boundary conditions}
We solve equations \eqref{eq:GS_eqn_04} and \eqref{eq:Fpsi_dimensionless_01} assuming an initially dipolar magnetic field
\begin{equation}
    \psi_{\rm d}(r,\theta) = \frac{\psi_\ast R_\ast \sin^2\theta}{r},
    \label{eq:dipole_initial_condition}
\end{equation}
where the magnetic field is anchored to the crust at $r=R_\ast$ before and after accretion. Following \citet{2004MNRAS.351..569P}, we employ the line-tying boundary condition at the stellar surface, $\psi(R_\ast,\theta) = \psi_{\rm d}(R_\ast,\theta)$, and the Dirichlet condition $\psi(r,0) = 0$ at the magnetic pole. At the equator, we enforce symmetry across the equatorial plane with the Neumann boundary condition $\partial\psi/\partial\theta = 0$, which ensures field lines are perpendicular to the equator at $\theta=\pi/2$.
We place the outer boundary at $r=R_{\rm max}$ sufficiently distant from the surface, such that all screening currents from the magnetic field distortion are contained within $r\ll R_{\rm max}$. In practice, we have $R_{\rm max} \gtrsim 10^5 x_0$ ($\tilde{x}_{\rm max} = 5\times10^4$). At the outer boundary, we follow \citet{2023MNRAS.526.2058R} in demanding that the magnetic dipole moment 
\begin{equation}
    |\mathbf{m}_{\rm d}| = \frac{3r^3}{4}\int_{-1}^{1}d(\cos\theta)\,\cos\theta B_r(r,\theta),
    \label{eq:dipole_moment_BC_01}
\end{equation}
is independent of $r$. The resulting boundary condition, as discussed by \citet{2023MNRAS.526.2058R}, becomes
\begin{equation}
    \frac{\partial\psi}{\partial r} = -\frac{\psi}{r},
    \label{eq:dipole_moment_BC_02}
\end{equation}
which is a Robin-type boundary condition to be applied alongside those above.
\subsection{Integral mass-flux constraint}\label{sec:app_b2_massflux_constraint}
In order to solve equations \eqref{eq:GS_eqn_04} and \eqref{eq:Fpsi_dimensionless_01}, we select an initial mass-flux distribution
\begin{equation}
    \frac{d\tilde{M}}{d\tilde{\psi}} = \frac{1}{2}\frac{\psi_0 b}{\psi_*}\frac{e^{-b\psi_0\tilde{\psi}/\psi_*}}{1 - e^{-b}}
    \label{eq:dmdpsi_02}
\end{equation}
 and an initial guess for the flux function, corresponding to a dipolar magnetic field, \(\psi^{(0)}(r,\theta) = \psi_*R_*\sin^2\theta\). In equation \eqref{eq:dmdpsi_02}, \(\psi_* = \psi(R_*, \pi/2)\) is the equatorial flux surface, \(\psi_{\rm a}\) is the flux surface that is connected to the inner edge of the accretion disk, such that accretion occurs on the polar cap over the range \(0 \leq \psi \leq \psi_{\rm a}\), and we define \(b = \psi_*/\psi_{\rm a}\). The coordinates of the contours of \(\psi\) are calculated using the Python library \texttt{contourpy}. Integrating along each contour according to equation \eqref{eq:Fpsi_dimensionless_01} yields \(\tilde{F}(\tilde{\psi})\). We choose \(N_c = N_r - 1\) contours in order to ensure the contours and grid points are spaced comparably, and thus roughly optimally, as validated by \citet{2004MNRAS.351..569P}. 
 
 To calculate the derivative \(\tilde{F}^{\prime}(\tilde{\psi})\), a univariate spline is fitted to \(\tilde F(\tilde \psi)\), using the \texttt{scipy} \citep{2020SciPy-NMeth} interpolation package's \texttt{UnivariateSpline} class, which can then be differentiated. This avoids numerical instabilities introduced by finite differencing. We find it to be more stable than the polynomial fit used by \citet{2004MNRAS.351..569P}. 
 
 At the equator, there are large gradients in $\psi$. Flux surfaces emerging from $r=R_\ast$ must bend sharply to become perpendicular to the equatorial plane due to the north-south symmetry. The steepness of the gradients hampers the convergence of our numerical scheme, unlike in \citet{2004MNRAS.351..569P}, where negligible mass is loaded onto flux tubes near the equator in the absence of cross-field mass transport. To promote convergence, we smooth $\tilde{F}(\tilde{\psi})$ by excluding arbitrarily the last three flux surfaces in the spline interpolation. More detailed modelling of the equatorial region lies outside the scope of this paper.
\subsection{Grad-Shafranov solver}\label{sec:app_b3_GS_solver_SOR}
The source term corresponding to the right-hand side of equation \eqref{eq:GS_eqn_04} is calculated after mapping \(\tilde F^{\prime}(\tilde \psi)\) to the grid using bilinear interpolation. Equation \eqref{eq:GS_eqn_04} is an elliptical partial differential equation and may be solved iteratively using successive over-relaxation to obtain the Gauss-Seidel iterate \(\tilde \psi_{\rm new}^{(0)}\). The next iterate is obtained by under-relaxation, viz. \(\tilde \psi^{(k+1)} = \Theta^{(k)}\tilde\psi^{(k)} + \left[1 - \Theta^{(k)}\right]\tilde\psi_{\rm new}^{(k)}\), with under-relaxation factor \(0\leq\Theta^{(k)}\leq 1\). We augment the relaxation scheme with Chebyshev acceleration, allowing the relaxation factor \(\Theta^{(k)}\)to vary to speed up convergence, by calculating the mean residual \(\delta^{(k)}\) at each iteration and comparing it to the previous iteration. The factor \(\Theta^{(k+1)}\) increases towards one if the residual \(\delta^{(k)}\) is larger than the residual at the previous iteration, \(\delta^{(k-1)}\), and decreases otherwise, according to
\begin{equation}
\Theta^{(k+1)} = 
    \begin{cases}
    \left(\Theta^{(k)}+1\right)/2 &\delta^{(k)} > \sigma \delta^{(k-1)}\\
    2\Theta^{(k)} - 1 &\delta^{(k)} < \sigma \delta^{(k-1)}
    \end{cases}
    \label{eq:update_sor_theta}
\end{equation}
In \eqref{eq:update_sor_theta}, $\sigma$ is a tolerance which prevents the relaxation parameter updating unless the residual changes between iterations by a factor $\sigma$. We choose $\sigma = 5$.
\subsection{Convergence}\label{sec:app_b7_convergence}
Convergence is validated for the solution of the Grad-Shafranov equation \eqref{eq:GS_eqn_04} in two ways. Firstly, the mean residual over the grid is calculated according to
\begin{equation}
    \left\langle\frac{\Delta\psi}{\psi}\right\rangle^{(k)} = \frac{1}{N_r N_\theta}\sum_{i,j}\frac{|\psi^{(k)}(x_i,\mu_j) - \psi^{(k-1)}(x_i,\mu_j)|}{|\psi^{(k)}(x_i,\mu_j)|}
    \label{eq:residual_01}
\end{equation}
and the iterative scheme continues until \(\left\langle\Delta\psi/\psi\right\rangle^{(k)} < \epsilon\) is satisfied, where we typically choose \(\epsilon = 10^{-3}\). Secondly, we track the total enclosed mass in the simulation domain, integrating the density \(\rho(x,\mu)\) over the grid at each iteration to calculate the enclosed mass \(M_{\rm check}\) according to
\begin{align}
    M_{\rm check} &= \int_V drd\theta d\phi\, \rho(r,\theta,\phi)r^2\sin\theta \\
    &= 2\pi x_0^3\int_0^{\tilde{x}_{\rm max}}d\tilde{x}\int_{0}^{1}d\tilde{\mu}\,(\tilde{x}+a)^2\rho(\tilde{x},\tilde{\mu})e^{-\tilde{x}}.
    \label{eq:m_check_continuous}
\end{align}
When converted to a discrete sum over grid points, and including a factor of two to account for both hemispheres, \eqref{eq:m_check_continuous} becomes
\begin{equation}
    M_{\rm check} = 4\pi x_0^3\sum_{i=0}^{N_{\tilde{x}}}\sum_{j=0}^{N_{\tilde{\mu}}}\Delta\tilde{x}\Delta\tilde{\mu}(\tilde{x}+a)^2\rho(\tilde{x},\tilde{\mu})e^{-\tilde{x}}.
    \label{eq:m_check_discrete}
\end{equation}
In practice, we require the enclosed mass to be within approximately five per cent of the accreted mass \(M_{\rm a}\) when solving the Grad-Shafranov equation, to ensure minimal mass leakage. 

\begin{figure}
\includegraphics[width=\columnwidth]{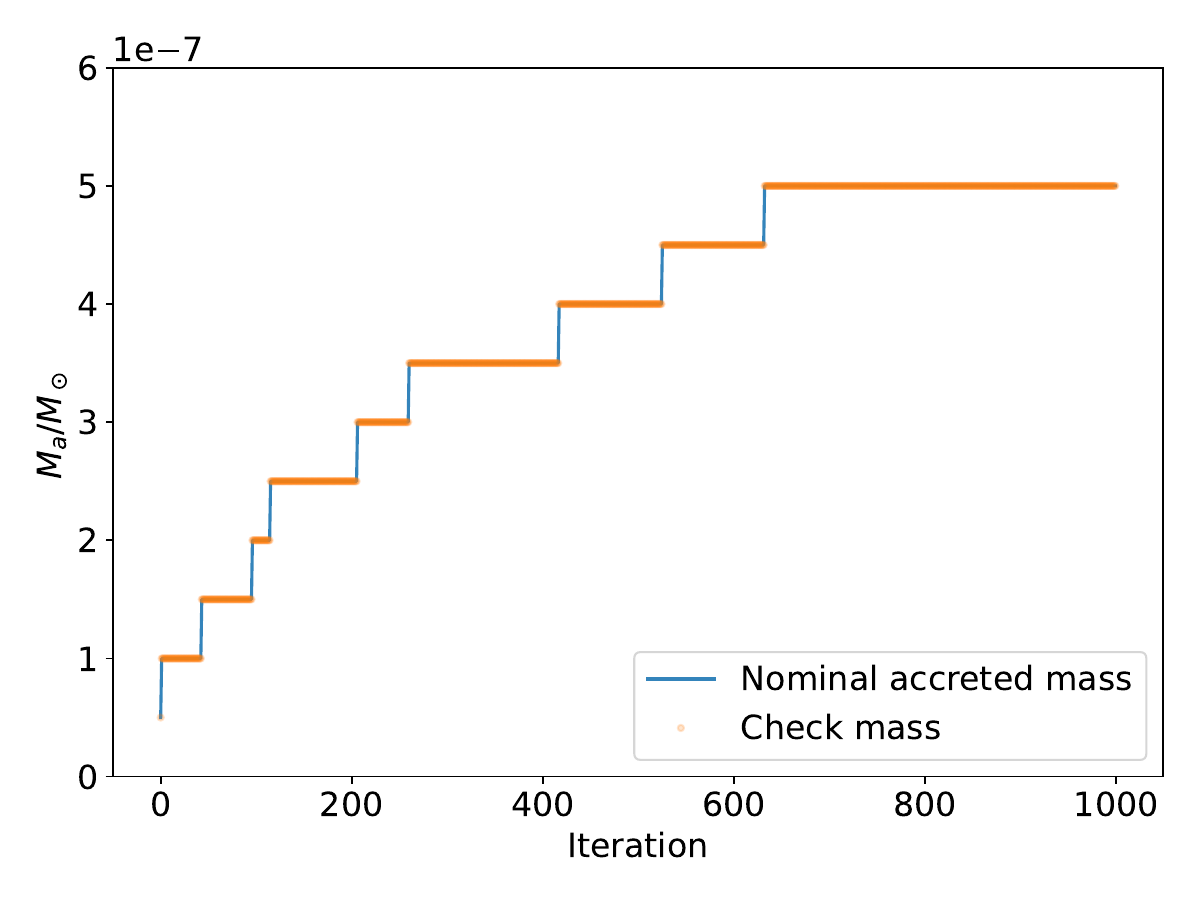}
\caption{Plot of the total mass in the domain (orange points) and the nominal mass (blue line) versus iteration number, during the check of  path dependence (Section \ref{sec:path_dependence}), demonstrating that mass does not leak appreciably; the orange points do not deviate from the blue curve during the assembly of the mountain.}
\label{fig:accreted_mass_b10_01}
\end{figure}

We verify that mass does not leak out of the simulation domain due to numerical errors, plotting $M_{\rm check}$ alongside the nominal mass added via \eqref{eq:update_dmdpsi_accretion_01} in Figure~\ref{fig:accreted_mass_b10_01}. The plot is flat during stages of mass transport, when no mass is added to the domain but shuffles between flux surfaces. There is negligible ($\leq0.1$ per cent) mass leakage throughout the iterative scheme. 

\subsection{Instability calculation}\label{sec:app_b4_instab_calc}
Having found the equilibrium solution, the locations of unstable points along each flux surface are calculated using equation \eqref{eq:delta_criterion_00}. The quantity \(\Delta\) may be written in terms of gradients in the \(\tilde{\mu}-\) and \(\tilde{x} -\)coordinates, such that equation~\eqref{eq:delta_criterion_01} becomes
\begin{equation}
\begin{aligned}[b]
\frac{d}{d\xi} \ln\left(\frac{p}{B^{\gamma_A}}\right)=\ &\frac{1}{x_0^2|\nabla\psi|}\Biggl[\frac{\partial \psi}{\partial \tilde{x}}\left(\frac{1}{p}\frac{\partial p}{\partial \tilde{x}} - \frac{\gamma_A}{B}\frac{\partial B}{\partial \tilde{x}}\right)\\
&+\frac{1 - \tilde{\mu}^2}{(\tilde{x} + a)^2}\frac{\partial \psi}{\partial \tilde{\mu}}\left(\frac{1}{p}\frac{\partial p}{\partial \tilde{\mu}} - \frac{\gamma_A}{B}\frac{\partial B}{\partial \tilde{\mu}}\right) \Biggr] .
\end{aligned}
\label{eq:delta_03}
\end{equation}
The code calculates \(g_\xi\) and \(\Delta\) at every point along each field line, and uses equation \eqref{eq:delta_criterion_00} to identify unstable points. 

To calculate \(\Delta_{\rm crit}\) we require a value for the characteristic length \(\ell\). We first note that \(\Delta_{\rm crit}\) is positive and typically small, \(\Delta_{\rm crit} \ll \Delta\). Hence it only affects the sign of equation~\eqref{eq:delta_criterion_00}, where \(\Delta\) is small and positive, which tends to occur in the upper atmosphere, rather than close to the stellar surface. In contrast, the unstable region within a few scale heights of the surface, where burial is concentrated, features larger values of \(\Delta\), and the effect of $\Delta_{\rm crit}$ is minimal there. Hence for each value of \(M_{\rm a}\), we first calculate \(g_\xi \Delta\) (i.e. without $\Delta_{\rm crit}$), and from the set of unstable flux surfaces that make up the resulting unstable region, we calculate \(\ell\) by calculating the average length of each field line that is unstable in that region.
\subsection{Cross-field mass transport}{\label{sec:app_b5_cfmf_calc}}
Once unstable flux surfaces are identified, the masses in the adjacent flux tubes are equalised, adjusting the mass-flux distribution by adding or subtracting the required \(\Delta M\) from each unstable flux tube according to \eqref{eq:update_dmdpsi_accretion_delta_m_00}. As described in Section \ref{subsec: cfmf}, when the unstable region extends over multiple adjacent flux surfaces, forming a contiguous unstable region, we calculate the total mass in every flux tube in the region and redistribute it equally between the same flux tubes. Specifically, we first calculate the cumulative mass up to the flux surface \(\psi\),
\begin{equation}
    M(\psi) = \int_{0}^{\psi}d\psi'\, \frac{dM}{d\psi'},
    \label{eq:flux tube_mass_03}
\end{equation}
then adjust the cumulative mass at each unstable flux surface by adding the appropriate increment \(\Delta M(\psi)\). We fit an interpolating spline using the \texttt{scipy.interpolate} package's \texttt{InterpolatedUnivariateSpline} class, and differentiate it to get \(dM/d\psi\) using the \texttt{numpy.gradient} function, a second-order central difference scheme \citep{harris2020array}. We track the total mass in the simulation domain to check mass conservation as discussed in Appendix \ref{sec:app_b7_convergence}.
\subsection{Accreting additional mass after nullifying the instability}\label{sec:app_b6_mass_accretion}
At the end of an iterative cycle for a given accreted mass \(M_{\rm a}^{(k)}\), once the unstable region disappears, more mass is added by modifying the mass-flux distribution according to \eqref{eq:update_dmdpsi_accretion_01}, by adding mass distributed in the same manner as the initial distribution, i.e. \eqref{eq:dmdpsi_02}. Expressed explicitly, the update step is
\begin{equation}
   \left(\frac{d\tilde{M}}{d\tilde{\psi}}\right)^{\rm (k+1)} = \left(\frac{d\tilde{M}}{d\tilde{\psi}}\right)^{\rm (adj)} + \frac{n_{\rm add}}{2}\frac{\psi_0 b}{\psi_*}\frac{e^{-b\psi_0\tilde{\psi}/\psi_*}}{1 - e^{-b}}.
   \label{eq:update_dmdpsi_accretion_02}
\end{equation}
The user-selected factor \(n_{\rm add}\) controls the rate of accretion for numerical stability purposes; it is a numerical parameter not a physical one. Typically we choose \(n_{\rm add} = 1\). The code first checks the total number of unstable points and only performs the update \eqref{eq:update_dmdpsi_accretion_02}, if the total number of unstable points is less than 0.1 per cent of the total number of points, which corresponds to the bottom left of \figureautorefname~\ref{fig:cfd-flowchart}.  To promote numerical stability, we exclude unstable surfaces within approximately 10 degrees of the equator when deciding whether to add extra mass according to \eqref{eq:update_dmdpsi_accretion_02}.
\section{Intermediate mass-flux distribution during quasistatic mountain assembly}\label{sec:app_C_dmdpsi_fit}
\begin{figure}
\centering
\includegraphics[width=\columnwidth]{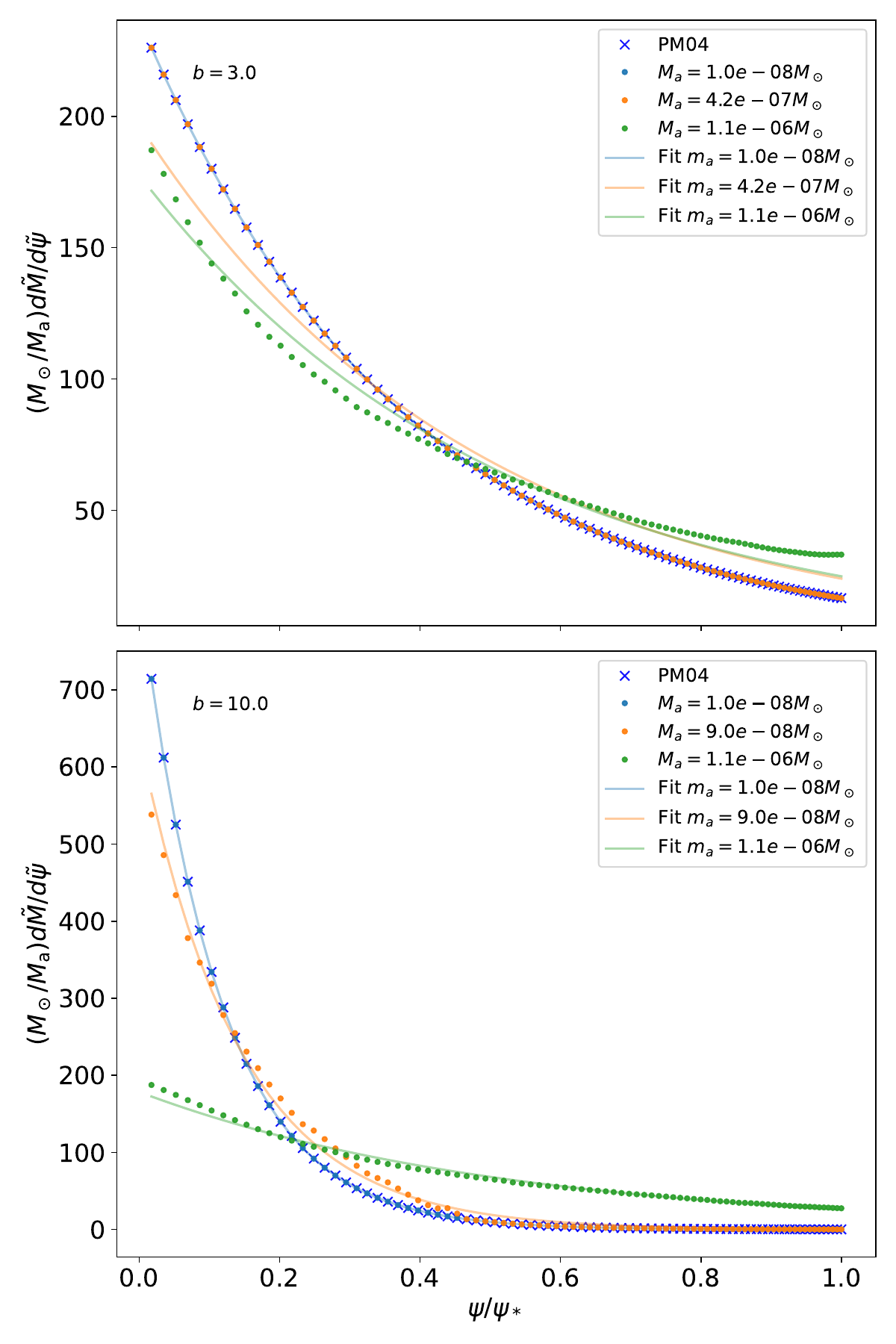}
\caption{Calibrating the parametric fit of the mass-flux distribution with cross-field transport at three main intermediate stages during quasistatic assembly. The first stage (blue points and curves) corresponds to the profile prior to the onset of the instability, corresponding to the profile of \citet{2004MNRAS.351..569P} (blue crosses). The second (orange points and curves) corresponds to the start of the instability, after the mountain is stabilised by a single round of mass transport according to the scheme described in Section \ref{sec:3}. The third (green points and curves) corresponds to the highest mass reached during the assembly sequence. The curves are fits of the form \eqref{eq:parameter_fit_sech}; the points are simulation outputs fitted by the curves. }
\label{fig:sech_fit_and_residuals}
\end{figure}

In this appendix, we discuss in more detail how the mass-flux distribution develops, as we iterate the quasistatic assembly process in Section \ref{sec:hydromag_structure_of_mtn} on the way to calculating the hydromagnetic structure of a representative mountain with $M_{\rm a} = 1\times 10^{-5} M_\odot$. In Figure \ref{fig:sech_fit_and_residuals}, we plot a subset of $dM/d\psi$ profiles selected from the calibration sequence depicted in Figure \ref{fig:dmdpsi_evolution_b10_01}. Specifically, in the top panel, we plot $dM/d\psi$ for $b=3$ at $M_{\rm a}'/M_\odot = 1\times10^{-8},\,4.2\times10^{-7}$, and $1.1\times10^{-6}$, whilst in the bottom panel we plot $dM/d\psi$ for $b=10$ at $M_{\rm a}'/M_\odot = 1\times10^{-8},\,9.0\times10^{-8}$, and $1.1\times10^{-6}$. The three plots in each panel represent three distinct stages in the iterative assembly of the mountain. The first profile (blue curve) is of the mountain before any unstable regions appear, when the profile matches the exponential form \eqref{eq:dmdpsi_pm04}. The second profile (orange curve) is an iterative snapshot after a period of accretion, at the moment immediately preceding the appearance of the first unstable region. The third profile (green) is an iterative snapshot at the end of quasistatic assembly. Upon comparing the three curves, one sees that mass migrates equatorwards. The position of the inflection point at $\psi/\psi_\ast \simeq 0.3$ depends weakly on $M_{\rm a}'$. This is especially apparent in the bottom panel of Figure \ref{fig:dmdpsi_evolution_b10_01}.

\begin{figure}
    \centering
    \includegraphics[width=\columnwidth]{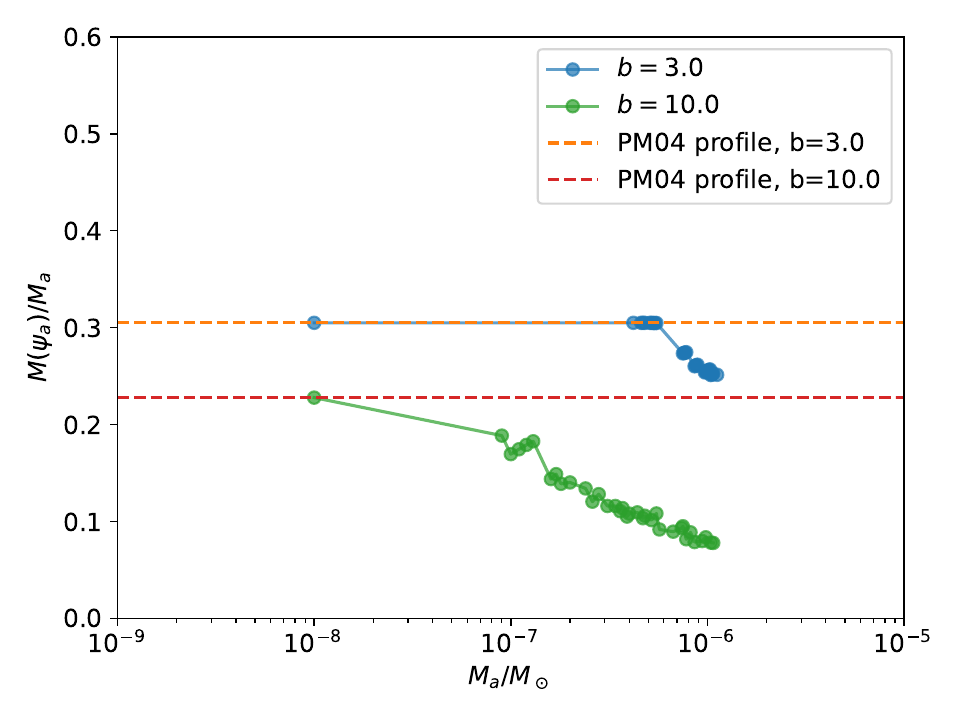}
    \caption{Equatorwards mass transport during quastistatic assembly of the mountain: fraction of the total accreted mass contained within the polar cap region $\psi \leq \psi_{\rm a} = \psi_\ast/b$, as a function of accreted mass for $b=3$ (blue points) and $b=10$ (green points). The dashed lines denote the mass contained in the polar cap prior to mass transport for the respective values of $b$. During accretion, cross-field transport reduces the total mass contained in the polar cap region. For $b=10$, over half the mass escapes from the cap for $M_{\rm a}/M_\odot\gtrsim5\times10^{-7}$.}
    \label{fig:mass_in_psi_a}
\end{figure}

Compared to the exponential form of $dM/d\psi$ \eqref{eq:dmdpsi_pm04} used by \citet{2004MNRAS.351..569P}, the intermediate $dM/d\psi$ profiles in \figureautorefname~\ref{fig:dmdpsi_evolution_b10_01} are shallower. We quantify this in Figure \ref{fig:mass_in_psi_a} by plotting the mass contained in the polar cap, $M(\psi \leq \psi_{\rm a})$, as a function of the accreted mass as the mountain is assembled and comparing it to $M(\psi \leq \psi_{\rm a})$ without mass transport. In the former case, we integrate the intermediate $dM/d\psi$ between $0\leq\psi\leq\psi_{\rm a}$ after mass transport stabilises the mountain; in the latter case, we integrate the original $dM/d\psi$ of equation \eqref{eq:dmdpsi_pm04}. For $b=10$, the original distribution without mass transport contains $0.23 M_{\rm a}$ within the polar cap. By contrast, for $M_{\rm a}'/M_\odot = 10^{-6}$, the quasistatically-assembled profile with mass transport contains $0.08 M_{\rm a}$ within the polar cap; $65$ per cent of the mass escapes equatorwards. Similarly, for $b=3$, the mass in the polar cap reduces from $0.31M_\odot$ to $0.26M_\odot$ between $M_{\rm a}'/M_\odot=4.2\times10^{-7}$ and $10^{-6}$, a reduction of $16$ per cent.

\section{Mountain on a newborn neutron star}
In this paper, we assume accretion occurs over long time-scales $M_{\rm a} / \dot{M}_{\rm a} \gtrsim 10^8 \, {\rm yr}$, characteristic of low mass X-ray binaries. One may ask whether magnetically supported mountains can also exist on newborn neutron stars, where one has $\dot{M}_{\rm a} \gtrsim 10^{-4}\, M_{\odot}\,\rm s^{-1}$ and $M_{\rm a} / \dot{M}_{\rm a} \sim 10^{3}\rm \,s$ \citep{2011ApJ...736..108P,2012ApJ...761...63P,2014ApJ...794..170M}. One such observationally motivated scenario occurs for protomagnetars accreting supernova fallback material, studied in detail by \citet{2014ApJ...794..170M}.

In the fallback scenario, a protomagnetar born with $B_\ast \gtrsim 10^{11}\,$T spins up to millisecond periods by accretion funnelled onto the magnetic poles. The accretion rate is high compared to a low mass X-ray binary. Fallback proceeds in two stages, $\dot M_{\rm early}\approx 10^{-3}\eta t^{1/2}\,M_{\odot}\,\rm s^{-1}$ and $\dot M_{\rm late}\approx 50 t^{-5/3}\,M_\odot\,\rm s^{-1}$, where $t$ is the time after core bounce, and $\eta$ is a dimensionless constant depending on the explosion energy $E_s$, with $0.2\leq\eta\leq 10$ for $0.3\leq E_s/10^{51}\,\rm erg \leq 1.2$. Overall it lasts $\sim 10^3\,$s, with the transition between early and late accretion occurring at $t\sim 10^2\,$s. The total accretion rate is $\dot M_a = (\dot M_{\rm early}^{-1} + \dot M_{\rm late}^{-1})^{-1}$, and exceeds the rate studied in this paper by $\gtrsim10\,$ orders of magnitude.

Despite the disparity in $\dot{M}_{\rm a}$, the mechanism of mountain building is the same: accreted material is funnelled to the surface at the magnetic poles, where it drags the magnetic field equatorwards by flux-freezing, and the same Grad-Shafranov equilibrium satisfying \eqref{eq:GS_eqn_03} and \eqref{eq:Fpsi_01} occurs. The magnetic mountain, passing through a quasistatic sequence of equilibria with increasing $M_{\rm a}$, possesses a magnetic dipole moment and mass quadrupole moment which scale roughly as $|\mathbf{m}_{\rm d}| = |\mathbf{m}_{\rm i}|(1 + M_{\rm a}/M_{\rm c})^{-1}$ and $\epsilon = (M_{\rm a}/M_{\odot})(1 + M_{\rm a}/M_{\rm c})^{-1}$ respectively, reaching $\epsilon = 5\times 10^{-4}$ and $|\mathbf{m}_{\rm d}| = 0.6|\mathbf{m}_{\rm i}|$ at $M_{\rm a} = 1.6M_c$ for the representative parameters of \citet{2014ApJ...794..170M}, and $M_{\rm c} = 3\times 10^{-3}(|\mathbf{m}_{\rm i}|/10^{23}\,\mathrm{T\,m^3})^2M_{\odot}$. This is substantially higher than the ellipticities calculated in this paper, where it is found that the ellipticity instead saturates to $\epsilon\approx10^{-5}$ at similar values of $M_{\rm a}/M_{\rm c}$. Interestingly, the magnetic dipole moment burial $|\mathbf{m}_{\rm d}|/|\mathbf{m}_{\rm i}| \approx 0.5$ at $M_{\rm a}/M_{\rm c}\approx 1.6$ is similar to \citet{2014ApJ...794..170M}, despite the different physical mechanisms producing these results.

As $|\mathbf{m}_{\rm d}|$ reduces due to burial, the Alfv\'en radius $r_{\rm A} \propto |\mathbf{m}_{\rm d}|^{4/7}$ falls below the co-rotation radius, and the propeller effect is suppressed, which enables mass to continue accreting onto the protomagnetar. The total accreted mass integrated over the fallback episode reaches $\sim 2.0$, $3.0$, and $4.3\,M_\odot$ for $\eta = 0.3$, $1.0$, and $3.0$ respectively. As a result, in most cases the burial of $|\mathbf{m}_{\rm d}|$ assists black hole formation on a timescale $\sim t_{\rm pk} = 85\eta^{-6/13}\,$s.

The central result of Section \ref{sec:astro_observables} that cross-field mass transport driven by the Schwarzschild instability causes $\epsilon$ to saturate at $\sim 10^{-5}$ and $|\mathbf{m}|/|\mathbf{m}_i|$ to plateau at $\sim0.46$ has direct implications in the protomagnetar scenario. The idealised flux-freezing constraint assumed by \citet{2014ApJ...794..170M} places an upper bound on $\epsilon$ and $|\mathbf{m}_{\rm d}|$. In this paper we show that these upper bounds are not reached in practice: cross-field mass transport redistributes mass equatorwards, limiting the growth of the mountain.

In the fallback scenario, the effect of the Schwarzschild instability would be even more pronounced. The instability is triggered after the accretion of $M_{\rm a}/M_{\odot}=9\times 10^{-8}$, a threshold reached $9\times10^{-4} \, {\rm s}$ after the onset of fallback. The instability and subsequent cross-field mass transport operate on the Alfv\'en timescale $\tau_{\rm A}\sim 6.5\,(B/10^{8}\,{\rm T})^{-1}(\rho/10^{18}\,{\rm kg\,m}^{-3})^{1/2}$\,s, which is much shorter than both the accretion timescale and the fallback duration. The mass transport scheme would therefore operate continuously throughout the fallback episode, limiting $\epsilon$ below the values calculated by \citet{2014ApJ...794..170M} and reducing the effectiveness of dipole burial. This has important implications for the predictions of black-hole formation by \citet{2014ApJ...794..170M}. A proper quantitative assessment involves combining the mass transport scheme in Section \ref{sec:3} with the spin-evolution framework of \citet{2014ApJ...794..170M} and is left for future work.

The gravitational wave signal from protomagnetars is predicted to be a transient burst, whose peak wave strain and burst duration are controlled primarily by whether the fallback-driven spin up stalls in response to the magnetocentrifugal or gravitational wave torque. This is in contrast to the persistent narrowband emission of gravitational waves from low-mass X-ray binaries such as those considered in this paper.\label{sec:app_E_newborn}
\section{Effect of stellar rotation}
Stellar rotation modifies the analysis in this paper in two ways, which are discussed briefly in this appendix. First, it modifies the momentum equation \eqref{eq:momentum_cons_01} by adding centrifugal and Coriolis forces. The centrifugal force density $\mathbf{F}_{\rm cent} = \rho\mathbf{\Omega}\times(\boldsymbol{\Omega}\times\boldsymbol{r})$ does not disappear in the magnetostatic limit as it is independent of $\boldsymbol{v}$. It is absorbed into an effective pressure $ p_{\rm eff} = p - (\rho\Omega^2 r^2\sin^2\theta)/2$. Equation \eqref{eq:momentum_cons_01} then takes the same form as the non-rotating case with $p_{\rm eff}$ replacing the thermal pressure. The equilibrium is therefore unchanged by rotation, but the thermal pressure is recovered via $p = p_{\rm eff} + (\rho\Omega^2 r^2\sin^2\theta)/2\neq p_{\rm eff}$. We may estimate the importance of the correction due to the centrifugal force by considering the ratio
\begin{equation}
    \frac{|\mathbf{\Omega}\times(\boldsymbol{\Omega}\times\boldsymbol{r})|}{|\nabla\Phi|} \sim \frac{\Omega^2R_\ast^3}{GM_\ast}.
    \label{eq:centrifual_to_grav_ratio_0}
\end{equation}
Upon taking characteristic values $\Omega \approx 6\times10^2 \rm\,rad\,s^{-1}$, $R_\ast = 10^4\, \rm m$, $M_\ast = 3\times10^{30}\rm \, kg$, we find $\Omega^2R_\ast^3/GM_\ast \sim 10^{-3}$. In addition, the mountain is concentrated in the polar cap region close to the stellar rotation axis, and so the rotational correction is suppressed by a factor $b^{1/2}$ proportional to the square root of the polar cap radius.

The Coriolis force density $\mathbf{F}_{\rm Cor} = 2\rho\mathbf{\Omega}\times\boldsymbol{v}$ depends on the fluid velocity $\boldsymbol{v}$. It vanishes when we seek equilibrium solutions. However, it does affect the growth time-scale of the Schwarzschild instability and the rate and direction of cross-field mass transport. With respect to the growth time-scale, we may estimate its importance relative to the magnetic Lorentz force $(\nabla\times\mathbf{B})\times\mathbf{B}/\mu_0$ by considering the ratio
\begin{equation}
    \frac{|2\rho\mathbf{\Omega}\times\boldsymbol{v}|}{|(\nabla\times\mathbf{B})\times\mathbf{B}/\mu_0|} \sim \frac{2\rho v \mu_0 \Omega L}{B^2}.
    \label{eq:coriolis_to_lorentz_ratio_0}
\end{equation}
In \eqref{eq:coriolis_to_lorentz_ratio_0}, $L$ is a characteristic hydromagnetic length scale, which we may estimate to equal the hydrostatic scale height $x_0 \approx 0.5\,\rm m$, we have $B = 10^8\,\rm T$, and $\rho\approx10^{13}\,\rm kg\,m^{-3}$ is the density in the vicinity of the unstable region at the onset of the instability at $M_{\rm a}/M_\odot \sim 10^{-7}$. With $v_A^2 = B^2/\mu_0\rho$, the ratio \eqref{eq:coriolis_to_lorentz_ratio_0} reduces to $2v\Omega x_0/v_A^2\leq 2 \Omega x_0/v_{\rm A}$ for $v \leq v_{\rm A}$. Inserting characteristic values, we find $v_A \sim 3\times 10^4\,\rm m\,s^{-1}$, and $2\Omega x_0/v_A \lesssim3\times10^{-2}$: that is, the rotational correction is small. 

With respect to cross-field mass transport, we may compare $v<v_{\rm A}$ with the characteristic Bohm diffusion speed $v_{\rm B}\simeq D_{\rm B}/x_0$, where the Bohm diffusion coefficient is given by $D_{\rm B} = kT/16eB$. With a conservative estimate for characteristic temperatures of $\sim10^{10}\,$K in the mountain \citep{2019MNRAS.484.1079S}, the ratio of the two velocities is then
\begin{equation}
    \frac{v_{\rm B}}{v_{\rm A}} = 4\times10^{-8}\left(\frac{B}{10^8\,\rm T}\right)^{-2}\left(\frac{\rho}{10^{13}\,\rm kg\,m^{-3}}\right)^{1/2}\left(\frac{T}{10^{10}\,\rm K}\right).
    \label{eq:compare_bohm_alfven_velocities_ratio_scalings}
\end{equation}
Stellar rotation affects other complicated, time-dependent phenomena which are relevant to mountain formation but lie outside the scope of this paper. One example is the shape of the polar cap, which can take a non-circular shape (e.g., a ring), when rotation is fast \citep{1976MNRAS.175..395B,2025MNRAS.541.3280Y}, and can migrate stochastically across the stellar surface \citep{2008MNRAS.386..673K,2013MNRAS.433.3048K,2015SSRv..191..339R}. Another example is how the magnetic field oscillates when it is plucked, either during the development of the Schwarzschild instability or by ongoing fluctuations in $\dot{M}_{\rm a}$ during accretion. The oscillation frequencies split rotationally, whether the spectrum is discrete \citep{2006ApJ...652..597P, 2008mgm..conf.1158V} or continuous \citep{2009MNRAS.395.1963V}.
\label{sec:app_D_coriolis}

\bsp	
\label{lastpage}
\end{document}